\documentclass[12pt]{article}
\usepackage{mathrsfs}
\usepackage{amssymb}
\usepackage{color}
\usepackage{amsmath}
\usepackage{ascmac}
\usepackage{amsthm}
\usepackage[dvips]{graphicx}
\usepackage{booktabs}

\newtheorem{thm}{Theorem}
\newtheorem{lem}{Lemma}

\newtheorem{prp}{Proposition}

\newtheorem{as}{Assumption}

\def\be{{\beta}}

\def\ep{{\varepsilon}}
\def\la{{\lambda}}

\def\bbe{{\text{\boldmath $\beta$}}}

\def\beh{{\widehat \be}}

\def\lah{{\hat \la}}

\def\bbeh{{\widehat \bbe}}

\def\b{{\text{\boldmath $b$}}}

\def\x{{\text{\boldmath $x$}}}
\def\y{{\text{\boldmath $y$}}}

\def\O{{\text{\boldmath $O$}}}

\def\R{{\text{\boldmath $R$}}}

\def\Nc{{\cal N}}

\def\et{{\it et\, al.}}

\def\Ah{\widehat{A}}
\def\pd{\partial}

\title{Parametric Transformed Fay-Herriot Model for Small Area Estimation}

\author{Shonosuke Sugasawa\thanks{Graduate School of Economics, University of Tokyo, E-Mail: shonosuke622@gmail.com}\ \   and Tatsuya Kubokawa\thanks{Faculty of Economics, University of Tokyo, E-Mail: tatsuya@e.u-tokyo.ac.jp}
\\
{\it\normalsize University of Tokyo}
}
\date{}

\begin{document}
\maketitle
\begin{abstract}
In this paper, we consider parametric transformed Fay-Herriot models, and clarify conditions on transformations under which the estimator of the transformation is consistent.
It is shown that the dual power transformation satisfies the conditions.
Based on asymptotic properties for estimators of parameters, we derive a second-order approximation of the prediction error of the empirical best linear unbiased predictors (EBLUP) and obtain a second-order unbiased estimator of the prediction error.
Finally, performances of the proposed procedures are investigated through simulation and empirical studies.

\par\vspace{4mm}
\end{abstract}

\section{Introduction}\label{sec:int}

The linear mixed models (LMM) with both random and fixed effects have been extensively and actively studied from both theoretical and applied aspects  in the literature.
As specific normal linear mixed models, the Fay-Herriot model (Fay and Herriot, 1979) and the nested error regression models (Battese, Harter and Fuller, 1988) have been used in small-area estimation (SAE), where direct estimates such as sample means for small areas have unacceptable estimation errors because sample sizes of small areas are small.
Then the model-based shrinkage methods such as the empirical best linear unbiased predictor (EBLUP) have been utilized for providing reliable estimates for small-areas with higher precisions by borrowing data in the surrounding areas.
For a good survey on SAE, see Ghosh and Rao (1994), Rao (2003) and Pfeffermann (2013).
Also, see Hall and Maiti (2006a,b), Chamber, \et (2014), Chaudhuri and Ghosh (2011) and Opsomer, \et (2008) for recent articles on parametric and nonparametric approaches to SAE.  

This paper is concerned with flexible modeling for analyzing positive data in SAE.
A standard transformation of positive $y$ is the logarithmic transformation $\log(y)$, and Slud and Maiti (2006) used this method in the Fay-Herriot model.
This approach may be reasonable when the distribution of positive observations is positively skewed.
However, the log-transformation is not always appropriate.
An alternative conventional method is the Box-Cox power transformation (Box and Cox, 1964) given by
\begin{align*}
h^{BC}(y,\la)=\begin{cases}
\ (y^{\la}-1)/\lambda, \ \ \ &\la\neq 0,\\
\ \log y, \ \ \ \ &\la=0.
\end{cases}
\end{align*}
However, it should be noted that $h^{BC}(y,\la)$ is truncated as $h^{BC}(y,\la)\geq -1/\la$ for $\la>0$ and $h^{BC}(y,\la)\leq -1/\la$ for $\la<0$.
Thus, the Box-Cox transformation is not necessarily compatible with the normality assumption. 
Another drawback of the Box-Cox transformation is that the maximum likelihood (ML) estimator of the transformation parameter $\la$ is not consistent.
This negative property discourages us from using the Box-Cox transformation in SAE, because EBLUP which plugs in the ML estimator of $\la$ does not converge to the best predictor or the Bayes estimator.

In Section \ref{sec:model}, we consider the parametric transformations and the corresponding transformed Fay-Herriot models which apply the transformed observations to the standard Fay-Herriot model.
In Section \ref{sec:est}, we derive sufficient conditions which guarantee consistency of estimators for the three unknown parameters of the transformation parameter, regression coefficients and variance of a random effect.
It is shown that the conditions for consistency are satisfied by the dual power transformation described in Section \ref{sec:model}, while the Box-Cox transformation does not satisfy the conditions.
The EBLUP which plugs in the consistent estimators is suggested.
The EEBLUP is a reasonable procedure, since it converges to the BLUP or the Bayes estimator.

Measuring uncertainty of the EBLUP is important in the context of SAE, and two approaches to this issue are known:  One is to evaluate the EBLUP in terms of the mean squared error (MSE) (see Das \et, 2004, Datta \et, 2005 and Prasad and Rao, 1990), and the other is to construct the confidence interval based on the EBLUP (see Chatterjee \et, 2008, Diao \et, 2014 and Yoshimori and Lahiri, 2014b). 
In Section \ref{sec:mse}, we derive a second-order approximation of the MSE of the EBLUP.
A second-order unbiased estimator of the MSE is also provided via the parametric bootstrap method.

In Section \ref{sec:sim}, we investigate finite-sample performances of the suggested procedures by simulation.
The suggested procedures are also examined through analysis of the data in the Survey of Family Income and Expenditure (SFIE) in Japan.
All the technical proofs are given in Appendix.

\section{Parametric Transformed Fay-Herriot Models}\label{sec:model}

Let $h(y, \la)$ be a monotone transformation from $\mathbb{R}_+$ to $\mathbb{R}$ for positive $y$, where $\mathbb{R}$ and $\mathbb{R}_+$ denote the sets of real numbers and positive real numbers, respectively.
It is noted that the transformation involves unknown parameter $\la$.
It is assumed that positive data $y_1, \ldots, y_m$ are available, where $y_i$ is an area-level data like a sample mean for the $i$-th small area.
For $i=1,\ldots,m$, assume that the transformed observation $h(y_i, \la)$ has a linear mixed model suggested by Fay and Herriot (1979), given by
\begin{equation}
\label{model}
h(y_i,\la)=\x_i'\bbe+v_i+\ep_i, 
\end{equation}
where $\x_i$ is a $p$-dimensional known vector, $\bbe$ is a $p$-dimensional unknown vector of regression coefficients, $v_i$ is a random effect associated with the area $i$ and $\ep_i$ is an error term. 
It is assumed that $v_i$, $\ep_i$, $i=1, \ldots, m$, are mutually independently distributed as $v_i\sim\Nc(0, A)$ and $\ep_i \sim \Nc(0,D_i)$, where $A$ is an unknown common variance and $D_1, \ldots, D_m$ are known variances of the error terms. 

When we use the Fay-Herriot model for analyzing real data, we need to estimate $D_1,\ldots,D_m$ before applying the model.
Fay and Herriot (1979) employed generalized variance function methods that use some external information in the survey.
For more explanation, see Hawala and Lahiri (2010).
In our analysis given in Section \ref{sec:data},  we estimate $D_i$ using data in the past ten years, where we need to incorporate the estimation of the transformation parameter in (\ref{model}).
The method for estimating $D_i$ in (\ref{model}) is given in Section \ref{sec:data}.
Thus, it should be noted that all the theory described in the paper are correct under the conditional model given the value $D_1,\ldots,D_m$.

In this paper, we want to consider a class of the transformations $h(y,\la)$ so that the ML estimator of $\la$ is consistent.
To this end, we begin by describing the conditions on $h(y,\la)$.
For notational convenience, let $h_{a_1a_2,\ldots,a_n}(y,\la)$ for $a_1,\ldots,a_n\in \{y,\la\}$ be the partial derivative of $h(y,\la)$.

\begin{as}
\label{as:1}{\rm 
The following are assumed for the transformation $h(y,\la)$:
\begin{itemize}
\item[(A.1)]
$h(y,\la)$ is an monotone function of $y$ ($y>0$) and its range is $\mathbb{R}$.

\item[(A.2)]
The partial derivatives$h_{y}(y,\la), \ h_{\la}(y,\la), \ h_{\la\la}(y,\la), \ h_{y\la}(y,\la)$ and $h_{y\la\la}(y,\la)$
exist and they are continuous.

\item[(A.3)] 
Transformation function $h(y,\la)$ satisfies the integrability conditions given by
\begin{align*}
E&\left[h^2(y,\la)h_{\la}(y,\la)^2\right]=O(1), \quad E\left[h_{\lambda}(y,\la)^2\right]=O(1)\\
&E\bigl[| h_{\lambda\lambda}(y,\la) |\bigr]=O(1), \quad E\Bigl[ \Bigl| \frac{d}{d\la}\Bigl(\frac{h_{y\lambda}(y,\la)}{h_y(y,\la)}\Bigr) \Bigr| \Bigr]=O(1),
\end{align*}
where $h(y,\la)$ is normally distributed.
\end{itemize}
}
\end{as}

Assumption (A.1) means that the transformation is a one-to-one and onto function from $\mathbb{R}_+$ to $\mathbb{R}$.
Clearly, (A.1) is not satisfied by the Box-Cox transformation, but by $\log(y)$.
Assumptions (A.2) and (A.3) will be used to show consistency of estimators of $\la$ and to evaluate asymptotically MSE of the EBLUP.

A useful transformation satisfying Assumption \ref{as:1} is the dual power transformation suggested by Yang (2006), given by
\begin{align}
\label{DPT}
h^{DP}(y,\la)=\begin{cases}
\ (y^{\la}-y^{-\la})/2\la, \ \ \ &\la>0,\\
\ \log y, \ \ \ \ &\la=0.
\end{cases}
\end{align}
This transformation will be used in simulation and empirical studies in Section \ref{sec:sim}.
It is noted that for $z=h^{DP}(y,\la)$, the inverse transformation is expressed as
$$
y=\Bigl(\la z+\sqrt{\la^2z^2+1}\Bigr)^{1/\la}
$$
for $\la> 0$, and $y=e^{z}$ for $\la=0$. 
It can be verified that $h^{DP}(y,\la)$ satisfies Assumption \ref{as:1}, where the proof will be given in Appendix.

\begin{prp}
\label{prop:1}
The dual power transformation (\ref{DPT}) satisfies Assumption \ref{as:1}.
\end{prp}

\section{ Consistent Estimators of Parameters}
\label{sec:est}

In this section, we derive consistent estimators of the parameters $\bbe$, $A$ and $\la$ in model (\ref{model}).
We first provide estimators $\Ah(\la)$ and $\bbeh(\la)$ of $A$ and $\bbe$, respectively, when $\la$ is fixed. 
We next derive an estimator $\lah$ by solving an equation for estimating $\la$, and then we get estimators $\Ah(\lah)$ and $\bbeh(\lah)$ by plugging in the estimator $\lah$.  

\subsection{Estimation of $\bbe$ and A given $\la$}

We begin by estimating $\bbe$ and $A$ when $\la$ is given.
In this case, the conventional procedures given in the literature for the Fay-Herriot model can be inherited to the transformed model.
Thus, for given $A$ and $\la$, the maximum likelihood (ML) or generalized least square (GLS) estimator of $\bbe$ is given by
\begin{equation}
\label{beta}
\bbeh(A,\la)=\Bigl\{\sum_{j=1}^m(A+D_j)^{-1}\x_j\x_j'\Bigr\}^{-1}\sum_{j=1}^m(A+D_j)^{-1}\x_jh(y_j,\la).
\end{equation}

Concerning estimation of $A$ given $\la$, we consider a class of estimators $\Ah(\la)$ satisfying the following assumption:

\begin{as}
\label{as:2}{\rm
The following are assumed for the estimator $\Ah(\la)$ of $A$:
\begin{itemize}
\item[(A.4)] \ $\Ah(\la)=A+O_p(m^{-1/2})$,

\item[(A.5)] \ $\partial \Ah(\la)/\partial \la=O_p(1)$,

\item[(A.6)] \ $\partial \Ah(\la)/\partial \la-E\bigl[\partial \Ah(\la)/\partial \la\bigr]={O}_p(m^{-1/2})$.

\end{itemize}
}
\end{as}

Assumption (A.4) implies that the estimator $\Ah(\la)$ is consistent.
Assumptions (A.5) and (A.6) will be used for approximating prediction errors of EBLUP. 
Let us define $\bbeh(\la)$ by
$$
\bbeh(\la) = \bbeh(\Ah(\la), \la),
$$
which is provided by substituting $\Ah(\la)$ into $\bbeh(A, \la)$ in (\ref{beta}).
Asymptotic properties of $\bbeh(\la)$ can be investigated under the following standard conditions on $D_i$ and $\x_i$.

\begin{as}
\label{as:3}{\rm 
The following are assumed for $D_i$ and $\x_i$:
\begin{itemize}
\item[(A.7)]
\ $m^{-1}\sum_{j=1}^m\x_j\x_j'$ converges to a positive definite matrix as $m\to\infty$.
\item[(A.8)]
\ There exist constants ${\underline D}$ and ${\overline D}$ such that ${\underline D} \leq D_i \leq {\overline D}$ for $i=1, \ldots, m$, and ${\underline D}$ and ${\overline D}$ are positive constants independent of $m$.
\end{itemize}
}
\end{as}

Since $\bbeh(A,\la) \sim \Nc_p(\bbe, \{\sum_{j=1}^m(A+D_j)^{-1}\x_j\x_j'\}^{-1})$, it is clear that $\bbeh(A,\la)$ is consistent and $\bbeh(A,\la)-\bbe=\O_p(m^{-1/2})$ under Assumption \ref{as:3}.
Asymptotic properties on $\bbeh(\la)=\bbeh(\Ah(\la), \la)$ are given in the following lemma which will be proved in Appendix.
This lemma will be used in Lemma \ref{lem:2} to show that some estimators of $A$ satisfy condition (A.6).

\begin{lem}
\label{lem:1}
Assume the conditions {\rm (A.4)} and {\rm (A.5)} in Assumption \ref{as:2} and Assumption \ref{as:3}.
Then it holds that $\bbeh(\la)-\bbe=\O_p(m^{-1/2})$ and 
$$
\partial \bbeh(\la)/\partial \la-E\Bigl[ \partial \bbeh(\Ah(\la)/\partial \la\Bigr]=\O_p(m^{-1/2}).
$$
\end{lem}

We here demonstrate that several estimators of $A$ suggested in the literature satisfy Assumption \ref{as:2} for fixed $\la$.
A simple moment estimator of $A$ due to Prasad and Rao (1990) is given by 
\begin{equation}
\label{PR}
\Ah_{PR}(\la)=(m-p)^{-1}\Bigl\{\sum_{j=1}^m(h(y_j,\la)-\x_j'\bbeh^{OLS})^2-\sum_{j=1}^m D_j\left\{1-\x_j'(\x'\x)^{-1}\x_j\Bigr\}\right\},
\end{equation}
where $\x=(\x_1,\ldots,\x_m)'$, and $\bbeh^{OLS}$ is the ordinary least squares (OLS) estimator 
$$
\bbeh^{OLS}=\left(\sum_{j=1}^m\x_j\x_j'\right)^{-1}\sum_{j=1}^m\x_jh(y_j,\la).
$$
Another moment estimator due to Fay and Herriot (1979), denoted by $\Ah_{FH}(\la)$, is given as a solution of the equation
\begin{equation}
\label{FH}
\sum_{j=1}^m (A+D_j)^{-1}\left\{h(y_j,\la)-\x_j'\bbeh(A,\la)\right\}^2=m-p.
\end{equation}
The maximum likelihood estimator (ML) of $A$, denoted by $\Ah_{ML}(\la)$, is obtained as a solution of the equation
\begin{equation}
\label{ML}
\sum_{j=1}^m(A+D_j)^{-2}\left\{h(y_j,\la)-\x_j'\bbeh(A,\la)\right\}^2=\sum_{j=1}^m(A+D_j)^{-1}.
\end{equation}
The restricted maximum likelihood estimator (REML) of $A$, denoted by $\Ah_{REML}(\la)$, is given as a solution of the equation
\begin{equation}
\label{REML}
\sum_{j=1}^m\frac{\left\{h(y_j,\la)-\x_j'\bbeh(A,\la)\right\}^2}{(A+D_j)^2}=\sum_{j=1}^m {1\over A+D_j}-\sum_{j=1}^m\frac{\x_j'\left\{\sum_{k=1}^m(A+D_k)^{-1}\x_k\x_k'\right\}^{-1}\x_j}{(A+D_j)^2}.
\end{equation}
Then, it can be verified that the above four estimators satisfy Assumption \ref{as:2}.
The proof will be given in Appendix.

\begin{lem}
\label{lem:2}
Under Assumption \ref{as:3}, the estimators $\Ah_{PR}(\la)$, $\Ah_{FH}(\la)$, $\Ah_{ML}(\la)$ and $\Ah_{REML}(\la)$ satisfy Assumption \ref{as:2}.
\end{lem}

\subsection{Estimation of transformation parameter $\la$}

We provide a consistent estimator of the transformation parameter $\la$.
For estimating $\la$, we use the log-likelihood function, which is expressed as
\begin{equation}
\label{like}
L(\la,A,\bbe)\propto-\frac12\sum_{j=1}^m\log(A+D_j)-\frac12\sum_{j=1}^m\frac{\left\{h(y,\la)-\x_i'\bbe\right\}^2}{A+D_j}+\sum_{j=1}^m\log h_y(y_j,\la).
\end{equation}
The derivative with respect to $\la$ is written as
\begin{align*}
F(\la,A,\bbe)\left(\equiv\frac{\partial L(\la,A,\bbe)}{\partial\la}\right)=\sum_{j=1}^m\frac{h_{y\lambda}(y_j,\la)}{h_y(y_j,\la)}-\sum_{j=1}^m(A+D_j)^{-1}\left\{h(y_j,\la)-\x_j\bbe\right\}h_{\lambda}(y_j,\la).
\end{align*}
Thus, we suggest estimator $\lah$ as a solution of the equation:
\begin{equation}
\label{lam}
F(\lah,\Ah(\lah),\bbeh(\lah))=0,
\end{equation}
where $\Ah(\la)$ is an estimator of $A$ satisfying Assumption \ref{as:2}.
Then, it is shown in the following lemma that the estimator derived from (\ref{lam}) is consistent.
The proof will be given in Appendix.

\begin{lem}
\label{lem:3}
Let $\lah$ be the solution of $(\ref{lam})$. 
Then, $\lah-\la=O_p(m^{-1/2})$ and $E[\lah-\la]=O(m^{-1})$ under Assumptions \ref{as:1}, \ref{as:2} and \ref{as:3}.
\end{lem}

\section{EBLUP and Evaluation of the Prediction Error}
\label{sec:mse}

We now provide the empirical best linear unbiased predictor (EBLUP) for small-area estimation and evaluate asymptotically the prediction error of EBLUP.
Since EBLUP includes the estimator of the transformation parameter in the transformed Fay-Herriot model, it is harder to evaluate the prediction error than in the non-transformed Fay-Herriot model.
To this end, the asymptotic results derived in the previous section are heavily used.

\subsection{EBLUP}

Consider the problem of predicting $\eta_i=\x_i'\bbe+v_i$, which is the conditional mean of the transformed data given $v_i$, namely, $E[h(y_i,\la)|v_i]$.
The best predictor of $\eta_i$ is given by
\begin{equation}
\label{B}
\hat{\eta}_i^B(\bbe,A,\la)=\x_i'\bbe+\frac{A}{A+D_i}\bigl\{ h(y_i,\la)-\x_i'\bbe\bigr\}.
\end{equation}
Since $\bbe$, $A$ and $\la$ are unknown, we use the estimators suggested in Section \ref{sec:est}.
Substituting $\bbeh(A,\la) $, given in (\ref{beta}), into $\hat{\eta}_i^B(\bbe,A,\la)$ yields the estimator
$$
\hat{\eta}_i^{EB0}(A,\la)=\x_i'\bbeh(A,\la)+A(A+D_i)^{-1}\bigl\{ h(y_i,\la)-\x_i'\bbeh(A,\la)\bigr\},
$$
which is the best linear unbiased predictor (BLUP) as a function of $h(y_i,\la)$, $i=1, \ldots, m$.
For the parameters $A$ and $\la$, we use the estimators $\Ah(\lah)$ and $\lah$ suggested in Section \ref{sec:est}.
Substituting those estimators into the BLUP, we get the empirical best linear unbiased predictor (EBLUP) 
\begin{equation}
\label{EB2}
\hat{\eta}_i^{EB}=\x_i'\bbeh(\lah)+\frac{\Ah(\lah)}{\Ah(\lah)+D_i}\bigl\{ h(y_i,\lah)-\x_i'\bbeh(\lah)\bigr\}.
\end{equation}

\subsection{Second-order approximation of the prediction error}

The prediction error of EBLUP is evaluated in terms of the mean squared error (MSE) of $\hat{\eta}_i^{EB}$ given by
$$
{\rm MSE}_i(A, \la)=E\bigl[ (\hat{\eta}^{EB}_i-\eta_i)^2\bigr], 
$$
for $i=1,\ldots, m$.
It is seen that the MSE can be decomposed as
\begin{align}
E\bigl[(\hat{\eta}^{EB}_i-\eta_i)^2\bigr] 
&=E\bigl[(\hat{\eta}_i^{EB}-\hat{\eta}_i^{B})^2\bigr]+E\bigl[ (\hat{\eta}_i^{B}-\eta_i)^2\bigr]\notag\\
&=E\bigl[ (\hat{\eta}_i^{EB}-\hat{\eta}_i^{EB1})^2\bigr] + 2E\bigl[ (\hat{\eta}_i^{EB}-\hat{\eta}_i^{EB1})(\hat{\eta}_i^{EB1}-\hat{\eta}_i^{B})\bigr] \notag\\
&\ \ \ \ \ +E\bigl[ (\hat{\eta}_i^{EB1}-\hat{\eta}_i^{B})^2\bigr] + E\bigl[ (\hat{\eta}_i^{B}-\eta_i)^2\bigr],
\label{mse1}
\end{align}
where
$$
\hat{\eta}_i^{EB1}=\x_i'\bbeh(\la)+\frac{\Ah(\la)}{\Ah(\la)+D_i}\bigl\{ h(y_i,\la)-\x_i'\bbeh(\la) \bigr\}.
$$
It is noted that the first two terms in the r.h.s. of (\ref{mse1}) are affected by estimation error of $\lah$, but the last two terms are not affected, namely, $E[(\hat{\eta}_i^{EB1}-\hat{\eta}_i^{B})^2]$ and $E[(\hat{\eta}_i^{B}-\eta_i)^2]$ do not depend on randomness of $\lah$.
Thus, it follows from the well-known result in small area estimation (Datta, Rao and Smith, 2005) that under Assumption \ref{as:3},
\begin{equation}
E[(\hat{\eta}_i^{EB1}-\hat{\eta}_i^{B})^2]+E[(\hat{\eta}_i^{B}-\eta_i)^2]=g_{1i}(A)+g_{2i}(A)+g_{3i}(A)+O(m^{-3/2}),
\label{g123}
\end{equation}
where $g_{1i}(A)=AD_i/(A+D_i)$, $g_{2i}(A)=D_i(A+D_i)^{-2}\x_i'\bigl(\sum_{j=1}^m\x_j\x_j'(A+D_j)^{-1}\bigr)^{-1}\x_i$ and $g_{3i}(A)=2^{-1}D_i(A+D_i)^{-2}{\rm Var}(\Ah)$.
Thus, we need to evaluate the first two terms.

Since $\lah-\la=O_p(m^{-1/2})$ given in Lemma \ref{lem:3}, the first term can be approximated as 
$$
E[(\hat{\eta}_i^{EB}-\hat{\eta}_i^{EB1})^2]
=E\Bigl[ (\lah-\la)^2 \Bigl(\frac{\partial}{\partial\la}\hat{\eta}_i^{EB1}\Bigr)^2\Bigr]+O(m^{-3/2}).
$$
To estimate this term, the following lemma is helpful.

\begin{lem}
\label{lem:5}
Under Assumptions \ref{as:1}, \ref{as:2} and \ref{as:3}, the derivative of $\hat{\eta}_i^{EB1}$ is approximated as
\begin{align*}
\frac{\partial}{\partial\la}\hat{\eta}_i^{EB1}=R_{1i}+O_p(m^{-1/2}), 
\end{align*}
where
\begin{align*}
R_{1i}=&
\frac{A}{A+D_i}h_{\lambda}(y_i,\la)+\frac{D_i}{A+D_i}\x_i'\Bigl(\sum_{j=1}^m\frac{\x_j\x_j'}{A+D_j}\Bigr)^{-1}\sum_{j=1}^m\frac{\x_j}{A+D_j}E[ h_{\lambda}(y_j,\la) ]\\
&+\frac{D_i}{(A+D_i)^2}\left\{h(y_i,\la)-\x_i'\bbe\right\}r(A),
\end{align*}
and $r(A)$ is a leading term of $E\bigl[ \partial\Ah(\la)/\partial \la \bigr]$.
\end{lem}
It follows from Lemma \ref{lem:5} that $E[(\hat{\eta}_i^{EB}-\hat{\eta}_i^{EB1})^2]
=g_{4i}(A,\la) +O(m^{-3/2})$, where
\begin{align}
g_{4i}(A,\la)=E\bigl[(\lah-\la)^2 R_{1i}^2\bigr].
\label{g4}
\end{align}
For specific estimators of $A$, we can calculate values of $r(A)$.
For $\Ah_{FH}(\la), \Ah_{ML}(\la)$ and $\Ah_{REML}(\la)$, the values of $r(A)$ are given by
$$
r(A)=\Bigl(\sum_{j=1}^m(A+D_j)^{-k}\Bigr)^{-1}\Bigl(\sum_{j=1}^m(A+D_j)^{-k}E\left[\left\{h(y_j,\la)-\x_j'\bbe\right\}h_{\lambda}(y_j,\la)\right]\Bigr),
$$
where $k=1$ corresponds to $\Ah_{FH}(\la)$, and $k=2$ corresponds to $\Ah_{ML}(\la)$ and $\Ah_{REML}(\la)$. 
For $\Ah_{PR}(\la)$, the value of $r(A)$ is given by
$$
r(A)=\frac2{m-p}\sum_{j=1}^mE\left[\left\{h(y_j,\la)-\x_j'\bbe\right\}h_{\lambda}(y_j,\la)\right].
$$

For the second term, note that $\lah-\la=O_p(m^{-1/2})$, $\Ah(\la)-A=O_p(m^{-1/2})$ and $\bbeh(\la)-\bbe=\O_p(m^{-1/2})$.
Then it follows from Lemma \ref{lem:5} that
\begin{align}
2E[&(\hat{\eta}_i^{EB}-\hat{\eta}_i^{EB1})(\hat{\eta}_i^{EB1}-\hat{\eta}_i^{B})]
\notag\\
=& 2E\Bigl[\Bigl(\frac{\partial}{\partial\la}\hat{\eta}_i^{EB1}\Bigr)(\lah-\la)\Bigl\{\Bigl(\frac{\partial\hat{\eta}_i^B}{\partial\bbe}\Bigr)'(\bbeh-\bbe)+\Bigl(\frac{\partial\hat{\eta}_i^B}{\partial A}\Bigr)(\Ah-A)\Bigr\}\Bigr]+O(m^{-3/2})\notag\\
=&2E\Bigl[(\lah-\la)R_{1i}\Bigl(\frac{\partial\hat{\eta}_i^B}{\partial \bbe}\Bigr)'(\bbeh-\bbe)\Bigr]
+2E\Bigl[ R_{1i}\Bigl(\frac{\partial\hat{\eta}_i^B}{\partial A}\Bigr)(\lah-\la)(\Ah-A)\Bigr]+O(m^{-3/2})\notag\\
=&g_{5i}(A,\la)+O(m^{-3/2}),
\label{g5}
\end{align}
where 
$$
g_{5i}(A,\la) =2E[(\lah-\la)R_{1i}\R_{2i}'(\bbeh-\bbe)] +2E[R_{1i}R_{3i}(\lah-\la)(\Ah-A)]
$$
for 
$$
\R_{2i}=\frac{\partial\hat{\eta}_i^B}{\partial\bbe}=\frac{D_i}{A+D_i}\x_i, \quad
R_{3i}=\frac{\partial\hat{\eta}_i^B}{\partial A}=\frac{D_i}{(A+D_i)^2}\left\{h(y_i,\la)-\x_i'\bbe\right\}.
$$
It is noted that $g_{4i}(A,\la)$ and $g_{5i}(A,\la)$ are of order $O(m^{-1})$ and that $g_{4i}(A,\la)$ and $g_{5i}(A,\la)$ generally cannot be expressed explicitly.
Combining the above calculations gives the following theorem.

\begin{thm}
\label{thm:1}
Under Assumptions \ref{as:1}, \ref{as:2} and \ref{as:3}, the prediction error of EBLUP given in $(\ref{EB2})$ is approximated as
\begin{align*}
{\rm MSE}_i=g_{1i}(A)+g_{2i}(A)+g_{3i}(A)+g_{4i}(A,\la)+g_{5i}(A,\la)+O(m^{-3/2}),
\end{align*}
where $g_{ki}$, $k=1,\ldots 5$ are defined in $(\ref{g123})$, $(\ref{g4})$ and $(\ref{g5})$.
\end{thm}

\subsection{Second-order unbiased estimator of the prediction error}

For practical applications, we need to estimate the mean squared error of EBLUP. 
Although $g_{4i}(A,\la)$ and $g_{5i}(A,\la)$ are not expressed explicitly, we can provide their estimators using the parametric bootstrap method.

Corresponding to model (\ref{model}), random variable $y_i^\ast$ can be generated as 
$$
y_i^\ast=h^{-1}(\x_i' \bbeh+v_i^\ast+\epsilon_i^\ast,\lah), \ \ \ \  i=1,\ldots,m
$$
for $\bbeh=\bbeh(\Ah(\lah),\lah)$, where $v_i^{\ast}$'s and $\varepsilon_i^{\ast}$'s are mutually independently distributed random errors such that $v_i^{\ast}|\y\sim \Nc(0,\Ah)$ and $\varepsilon_i^{\ast}\sim \Nc(0,D_i)$ for $\y=(y_1, \ldots, y_m)$. 
The estimators $\lah^{\ast}$, $\bbeh^{\ast}$ and $\Ah^{\ast}$ can be obtained from $y_i^{\ast}$'s by using the same manners as used in $\lah$, $\bbeh$ and $\Ah$. 

Since $g_{2i}(A)+g_{3i}(A)=O(m^{-1})$, it is seen that $g_{2i}(\Ah)+g_{3i}(\Ah)$ is a second order unbiased estimator of $g_{2i}(A)+g_{3i}(A)$, namely $E[g_{2i}(\Ah)+g_{3i}(\Ah)]=g_{2i}(A)+g_{3i}(A)+O(m^{-3/2})$. 

For estimation of $g_{1i}(A)$, $g_{1i}(\Ah)$ has a second-order bias, since $g_{1i}(A)=O(1)$.
Thus, we need to correct the bias up to second order. 
By the Taylor series expansion of $g_{1i}(\Ah(\lah))$, 
\begin{align*}
E\bigl[g_{1i}(\Ah(\lah))\bigr]
&=E\Bigl[g_{1i}(A)+\{\Ah(\lah)-A\}\frac{d}{dA}g_{1i}(A)\Bigr]+O(m^{-1})\\
&=g_{1i}(A)+E\bigl[\Ah(\lah)-A\bigr]\frac{D_i^2}{(A+D_i)^2}+O(m^{-1}),
\end{align*}
and that
\begin{align*}
\Ah(\lah)-A
&=\Ah(\la)-A+(\lah-\la)\frac{\partial}{\partial\la}\Ah(\la)+O_p(m^{-1})\\
&=(\lah-\la)\Bigl\{ \frac{\partial}{\partial\la}\Ah(\la)-E\Bigl[ \frac{\partial}{\partial\la}\Ah(\la)\Bigr] \Bigr\} +(\lah-\la)E\Bigl[\frac{\partial}{\partial\la}\Ah(\la)\Bigr]+O_p(m^{-1}).
\end{align*}
Then it follows from Assumption \ref{as:2} and Lemma \ref{lem:3} that $E\bigl[ \Ah(\lah)-A \bigr]=O(m^{-1})$, which implies that
$$
E\bigl[ g_{1i}(\Ah(\lah)) \bigr] =g_{1i}(A)+b_i(A,\la)+O(m^{-3/2}),
$$
where $b_i(A,\la)$ is a bias with order $O(m^{-1})$. 
Hence, based on the parametric bootstrap, we get a second-order unbiased estimator of $g_{1i}(\Ah(\lah))$ given by
\begin{equation}
\label{BPg1i}
\overline{g_{1i}}(\Ah,\lah)= 2g_{1i}(\Ah(\lah)) - E^{\ast}\bigl[g_{1i}(\Ah^{\ast})|\y\bigr].
\end{equation}
In fact, it can be verified that $E[ \overline{g_{1i}}(\Ah,\lah)]= g_{1i}(A) + O(m^{-3/2})$, since $E^{\ast}[g_{1i}(\Ah^{\ast})|\b{y}]=g_{1i}(\Ah(\lah))+b_i(\Ah(\lah),\lah)+O_p(m^{-3/2})$.

For $g_{4i}(A,\la)$ and $g_{5i}(A,\la)$, their estimators based on the parametric bootstrap are given by
\begin{align*}
\overline{g_{4i}}(\Ah,\lah)
=& E_{\ast}\bigl[ (\hat{\eta}_i^{EB\ast}-\hat{\eta}_i^{EB1\ast})^2\bigr|\y],
\\
\overline{g_{5i}}(\Ah,\lah)
=& 
2E_{\ast}\bigl[ (\hat{\eta}_i^{EB\ast}-\hat{\eta}_i^{EB1\ast})(\hat{\eta}_i^{EB1\ast}-\hat{\eta}_i^{B\ast})\bigr|\y], 
\end{align*}
where 
\begin{align*}
\hat{\eta}_i^{B\ast}=&\x_i'\bbeh(\lah)+\frac{\Ah(\lah)}{\Ah(\lah)+D_i}\bigl\{ h(y_i^\ast,\lah)-\x_i'\bbeh(\lah)\bigr\},
\\
\hat{\eta}_i^{EB1\ast}=&\x_i'\bbeh^\ast(\lah)+\frac{\Ah^\ast(\lah)}{\Ah^*(\lah)+D_i}\bigl\{ h(y_i^\ast,\lah)-\x_i'\bbeh^*(\lah)\bigr\},
\\
\hat{\eta}_i^{EB\ast}=&\x_i'\bbeh^\ast(\lah^\ast)+\frac{\Ah^\ast(\lah^\ast)}{\Ah^\ast(\lah^\ast)+D_i}\bigl\{ h(y_i^\ast,\lah^\ast)-\x_i'\bbeh^\ast(\lah^\ast)\bigr\}.
\end{align*}

\medskip
Combining the above estimators yields the estimator of ${\rm MSE}_i$ given by
\begin{equation}
\label{MSEest}
\widehat{{\rm MSE}_i}^{\ast}=\overline{g_{1i}}(\Ah,\lah)+g_{2i}(\Ah)+g_{3i}(\Ah)+\overline{g_{4i}}(\Ah,\lah)+\overline{g_{5i}}(\Ah,\lah).
\end{equation}

\begin{thm}
\label{thm:2}
Under Assumptions \ref{as:1}, \ref{as:2} and \ref{as:3}, $\widehat{{\rm MSE}_i}^{\ast}$ is a second order unbiased estimator of MSE$_i$, that is 
$$
E[\widehat{{\rm MSE}_i}^{\ast}]={\rm MSE}_i+O\left(m^{-3/2}\right).
$$
\end{thm}

\section{Simulation and Empirical Studies}
\label{sec:sim}

In this section, we investigate finite-sample performances of estimators of the parameters, MSE of EBLUP and estimators of MSE through simulation experiments.
We also apply the suggested procedures to the data in the Survey of Family Income and Expenditure (SFIE) in Japan.

\subsection{Finite sample behaviors of estimators}

We first investigate finite sample performances of the proposed estimators in the model
$$
\frac{y_i^{\la}-y_i^{-\la}}{2\la}=\beta_0+\beta_1x_i+v_i+\ep_i, \quad i=1, \ldots, m.
$$ 
We generate covariates $x_i$ from $\Nc(0,1)$, and fix them through the simulation runs.
Let $\beta_1=0.5$, $\beta_2=1$, $A=0.4$, $\lambda=0.6$ and $m=30$ . 
In the simulation experiments, we generate 10,000 data sets of $y_i=h^{-1}(\beta_0+\beta_1x_i+v_i+\varepsilon_i,\lambda)$ for $i=1,\ldots,m$ to investigate performances of the estimators. 
The random effect $v_i$ is generated from $\Nc(0,0.4)$ with $A=0.4$, and the sampling error $\ep_i$ is generated from $\Nc(0,D_i)$.
For $D_i$'s, we treat the three patterns:
$$
{\rm (a)}\ 0.1, 0.2, 0.3, 0.4, 0.5;\ {\rm (b)}\ 0.1, 0.3, 0.5, 0.8, 1.0;\ {\rm (c)}\ 0.1, 0.4, 0.7, 1.1, 1.5.
$$ 
There are five groups $G_1,\ldots,G_5$ and six small areas in each group. 
The error variance $D_i$ is common in the same group.

For estimation of $A$, we use four methods of the maximum likelihood estimator (ML), restricted maximum likelihood estimator (REML), Prasad--Rao estimator (PR) and Fay--Herriot estimator (FH).
We also apply the log-transformed model for the simulated data, which corresponds to the case of $\lambda=0$ in the dual power transformation. 
For estimation $A$, $\beta_1$ and $\beta_2$ in the log-transformed model, we use the maximum likelihood method.

The average values of estimates  and standard errors of $\lambda$, $A$, $\beta_1$ and $\beta_2$ are reported in Table \ref{est}. 

It is observed that the estimates of $A$ in the logarithmic transformed case tend to underestimate $A$ and their performances are not as good as those in the parametric transformed case. 
Comparing the estimating method for $A$, we can see that the REML method gives the estimates closer to the true value of $A$ than the other methods. 

Recently, Li and Lahiri (2010) and Yoshimori and Lahiri (2014a) pointed out that zero estimates for $A$ in the Fay-Heriot model is not preferable since zero estimates for $A$ mean that resulting EBLUP estimates are over-shrunk to the regression estimator. 
Then, we calculated the percentage of zero estimates of $A$ based on $10,000$ simulation runs for various values of $\la$. 
The result is given in Figure 1 for pattern (a), (b) and (c). 
It is observed that the percentage in the log-transformation increases as $\la$ increases, so that it is better to use the parametric transformation for avoiding zero estimates for $A$.

Finally, we investigate robustness of the proposed estimators. 
Following Lahiri and Rao (1995), we considered two different distributions for the $v_i$'s, namely double exponential and location exponential, which have mean zero and variance $A=0.4$. 
The sampling error, $\ep_i$, was generated from $N(0,D_i)$ for $D_i$ specified by patterns (a)--(c). 
Since the simulation results of $\beta_1$ and $\beta_2$ are not very different from the result given in Table \ref{est}, we report average values and standard errors of estimators of $\la$ and $A$ for patterns (a) and (c) in Table \ref{robest}. 
Comparing these values with the corresponding average values given in Table \ref{est}, we note that the estimates of both $A$ and $\la$ in the double-exponential case perform as well as in the normal case. 
However, in the location-exponential case, the estimates of $A$ and $\la$ are more biased than both normal and double-exponential cases.
This may come from skewness of underlying distributions, since the location exponential is a skewed distribution, but the normal and the double-exponential are symmetric distributions.

\begin{table}[!htb]
\caption{ Average Values of Estimators of $A$ and $\lambda$ for $m=30$, $\beta_1=0.5,\beta_2=1$, $A=0.4$, $\la=0.6$, $D_i$-patterns (a), (b) and (c). (The standard erros are given in parentheses.)}
\begin{center}
\begin{tabular}{cccccccccccccc}
\toprule
&&\multicolumn{4}{c}{Pattern (a)} && \multicolumn{4}{c}{Pattern (b)}   \\ \midrule
Estimator of $A$ &&$\la$ & $A$ & $\beta_1$&$\beta_2$ && $\la$ & $A$ & $\beta_1$&$\beta_2$ \\ \midrule
ML&& 0.65&  0.46 & 0.53 &  1.06  &&   0.68  & 0.46  & 0.53 & 1.09 \\   
    &&(0.25)&(0.36)&(0.18)&(0.21)&&  (0.22)&(0.37)&(0.19)&(0.29)\\
REML&&0.63&  0.39 & 0.52   & 1.05   &&    0.67 & 0.41 & 0.53 & 1.07 \\
        &&(0.24)&(0.28)&(0.18)&(0.21)&&(0.27)&(0.33)&(0.20)&(0.29) \\
FH   &&0.67 & 0.47 &0.53 &1.08 &&0.66& 0.44 & 0.53 &1.08 \\
       &&(0.28) &(0.37) &(0.18)&(0.24)&&(0.22)&(0.39)&(0.20)&(0.26)\\
PR    &&0.67 &0.50 &0.54 &1.07  &&0.66 &0.55 &0.53 &1.08\\
       &&(0.23)& (0.35) &(0.19) &(0.23) &&(0.21) &(0.48)&(0.20)&(0.24)\\
log   &&---&0.16 & 0.42 & 0.84 &&  ---& 0.19 & 0.43 & 0.82\\
        &&--- &(0.10)&(0.12)&(0.09) &&  ---&(0.12)&(0.15)&(0.17) \\  \midrule
\end{tabular}

\bigskip
\begin{tabular}{cccccccccccccc}
\toprule
&& \multicolumn{4}{c}{Pattern (c)}  \\ \midrule
Estimator of $A$ &&$\la$ & $A$ & $\beta_1$&$\beta_2$ \\ \midrule
ML && 0.65  & 0.47 & 0.53 & 1.06&   \\   
    &&(0.21) &(0.49)&(0.22)&(0.23)  \\
REML&& 0.66& 0.39& 0.53& 1.05&    \\
        &&(0.22) &(0.34) &(0.20)&(0.24)  \\ 
FH    &&0.65  &0.44 &0.53 &1.06\\
        &&(0.21)&(0.46)&(0.21)&(0.24)\\
PR    &&0.62 & 0.52 & 0.52 & 1.04 \\
       &&(0.26)&(0.51)&(0.21)&(0.29)\\
log   &&  ---&0.11& 0.41&  0.82\\
       & &  ---&(0.10)&(0.14)&(0.09) \\  \midrule
\end{tabular}

\label{est}
\end{center}
\end{table}

\begin{figure}
\centering
\includegraphics[width=5.2cm,clip]{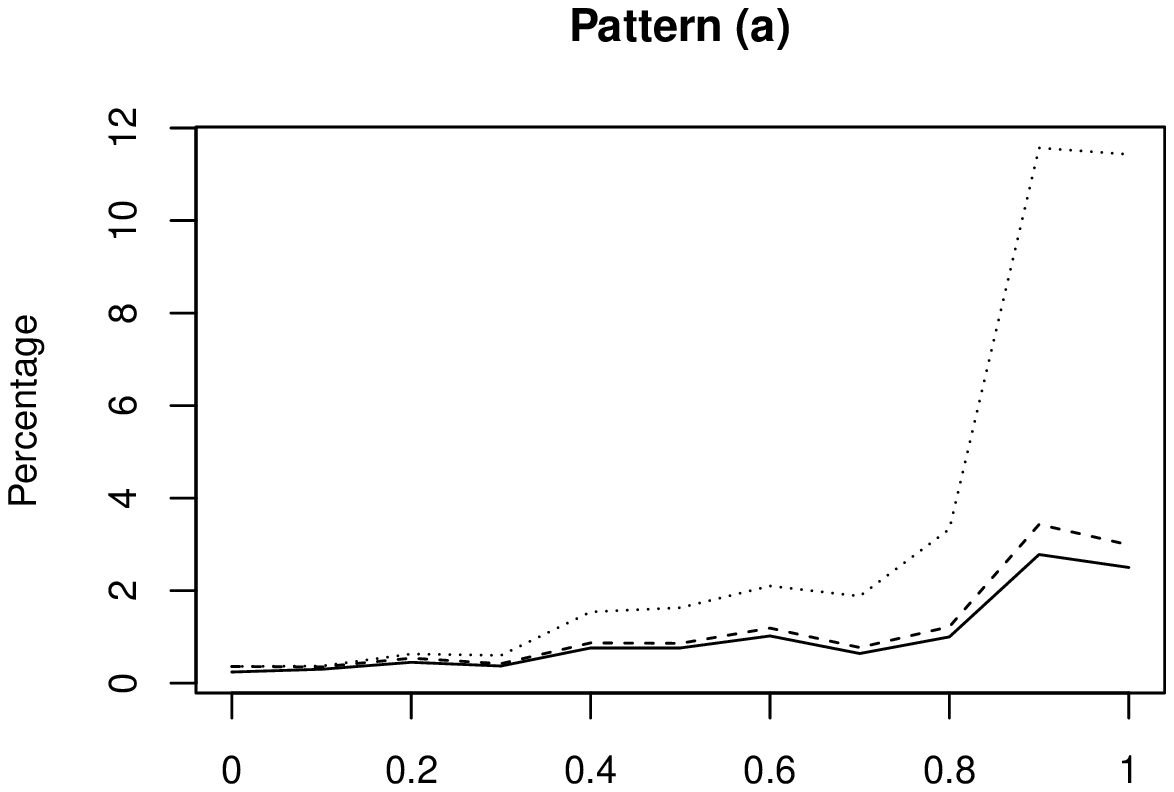}\ \ 
\includegraphics[width=5.2cm,clip]{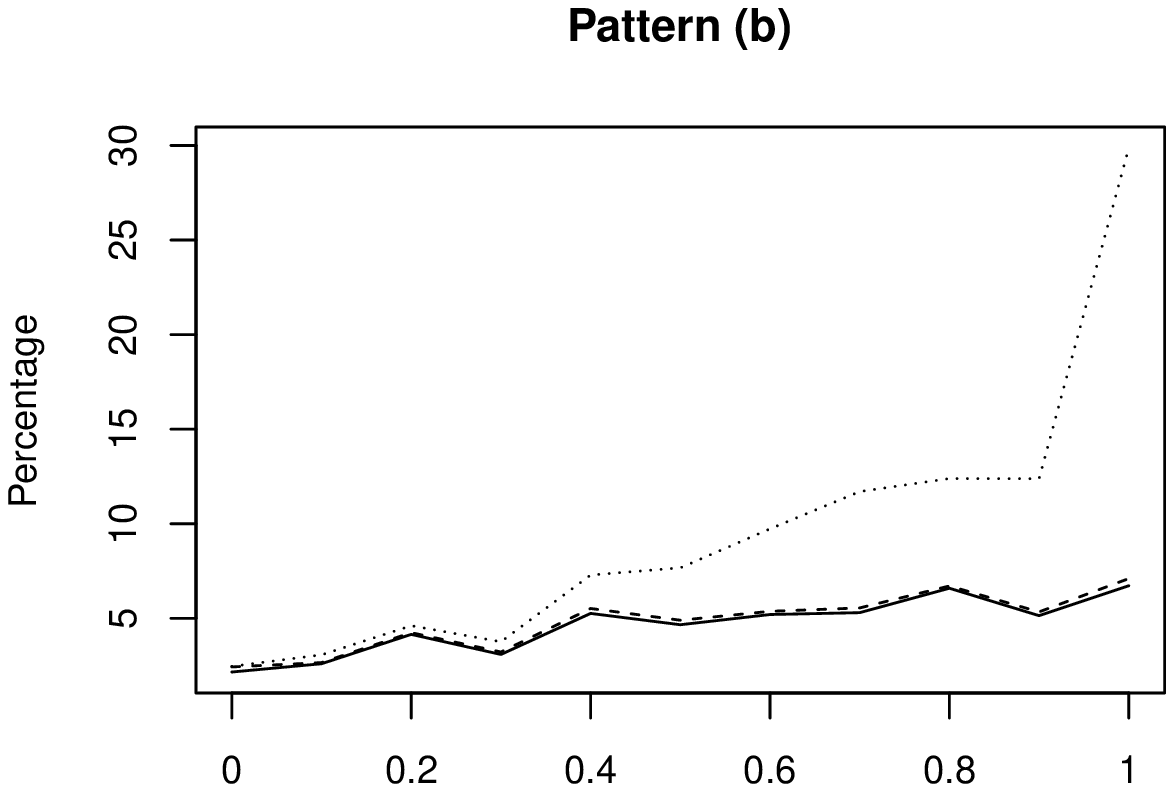}\ \ 
\includegraphics[width=5.2cm,clip]{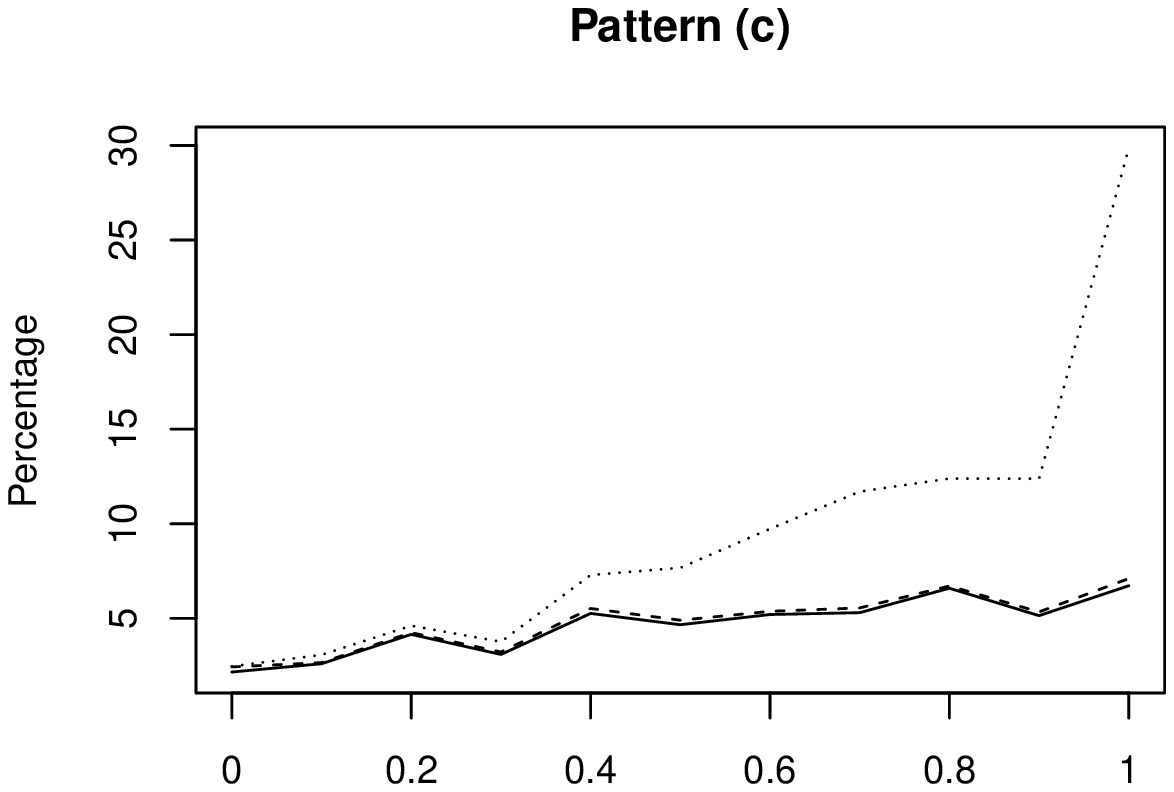}
\caption{Percentage of Zero Estimates of A in Pattern (a) and (c). (The horizontal axis indicates values of $\la$. The solid line corresponds to ML method, the dashed line to REML method and the dotted line to log-transformed model.)}
\end{figure}

\begin{table}[!htb]
\caption{ Average Values and Standard Errors of Estimators of $\beta_1,\beta_2,A$ and $\lambda$ for $m=30$, $\beta_1=0.5,\beta_2=1$, $A=0.4$, $\la=0.6$, $D_i$-patterns (a), (b) and (c), and for Double-exponential and Location-exponential Random Effects Distributions. (The standard erros are given in parentheses.)}
\begin{center}
\begin{tabular}{cccccccccccccc}
\toprule
&&\multicolumn{8}{c}{Double-exponential} \\
&&\multicolumn{2}{c}{Pattern (a)} && \multicolumn{2}{c}{Pattern (b)} &&\multicolumn{2}{c}{Pattern (c)} \\   \midrule
Estimator of $A$ && $\la$ & $A$ && $\la$ & $A$ &&  $\la$ & $A$ \\ \midrule
ML&& 0.60 & 0.41 && 0.63 &0.40 &&  0.64& 0.41\\
   &&(0.32)&(0.34)&&(0.26) & (0.36)&&  (0.21)&(0.41) \\
REML&&0.56& 0.36 && 0.61 & 0.36 &&0.63 & 0.36  \\
&&(0.32) &(0.29) &&(0.25) &(0.31)&& (0.20)&(0.36) \\
\midrule
\end{tabular}
\begin{tabular}{cccccccccccccc}
\toprule
&&\multicolumn{8}{c}{Location-exponential}  \\
&&\multicolumn{2}{c}{Pattern (a)} && \multicolumn{2}{c}{Pattern (b)} &&\multicolumn{2}{c}{Pattern (c)} \\   \midrule
Estimator of $A$ && $\la$ & $A$ && $\la$ & $A$ &&  $\la$ & $A$\\ \midrule
ML&&0.45 & 0.27  && 0.56 & 0.32 &&0.57 & 0.30  \\
   &&(0.33)&(0.22) && (0.27)&(0.29)&&(0.22) &(0.31)\\
REML&& 0.42&0.24 &&0.55 & 0.29 && 0.55& 0.27\\
&&(0.32)&(0.19) &&(0.26) & (0.26) && (0.21)&(0.27)\\
\midrule
\end{tabular}
\label{robest}
\end{center}
\end{table}

\subsection{Numerical properties of MSE and the estimators}

We next investigate MSE of EBLUP $\hat{\eta}_i^{EB}$ and performances of estimators of MSE.
The simulation experiments are implemented in the similar framework as treated in Datta $\et$ (2005).
Since MSE is location invariant, we consider the model (\ref{model}) without covariates namely $\x_i\bbe=\mu$, where the transformation function is the dual power transformation.
Let $\mu=0$ and $A=1$.
Let $\{Y_i^{(s)},i=1,\ldots,m\}$ be simulated data in the $s$-th replication for $s=1,\ldots$, $100,000 (=S)$.
Let $\hat{\eta}_i^{EB(s)}$ be EBLUP and let $\hat{\eta}_i^{B(s)}$ be the best predictor for the $s$-th replication. 
Also let $h(y_i^{(s)}, \lah^{(s)})$ be the direct predictor for the $s$-th replication.
Then the true values of MSE of EBLUP and the direct predictor $h(y_i,\lah)$ can be numerically obtained by 
$$
{\rm MSE}(\hat{\eta}_i^{EB}) \approx S^{-1}\sum_{s=1}^S\left(\hat{\eta}_i^{EB(s)}-\hat{\eta}_i^{B(s)}\right)^2+AD_i/(A+D_i),
$$
$$
{\rm MSE}(h(y_i,\lah)) \approx S^{-1}\sum_{s=1}^S\left(h(y_i^{(s)}, \lah^{(s)})-\hat{\eta}_i^{B(s)}\right)^2+AD_i/(A+D_i),
$$
and their averages over six small areas within group $G_i$ are denoted by ${\rm MSE}_{{\rm EBLUP}}(G_i)$ and ${\rm MSE}_{{\rm DP}}(G_i)$ for $i=1, \ldots, 5$.
The true values of ${\rm MSE}_{{\rm EBLUP}}(G_i)$ and the percentage relative gain in MSE defined by $100\times \bigl\{{\rm MSE}_{{\rm DP}}(G_i) - {\rm MSE}_{{\rm EBLUP}}(G_i)\bigr\}/{\rm MSE}_{{\rm DP}}(G_i) $ are reported in Table \ref{truemse}, where values of the percentage relative gain in MSE are given in parentheses. 
It is noted that EBLUP is a shrinkage predictor and $h(y_i,\lah)$ is the non-shrinkage direct predictor.
Thus, large values of the relative gain in MSE mean that the improvements of EBLUP over the direct predictor are large.
Table \ref{truemse} reveals that for all groups, the prediction error of EBLUP is smaller than that of the direct predictor.
Especially, the improvement of EBLUP seems significant in $G_3$, $G_4$ and $G_5$. 
This implies that EBLUP works well still in the transformed Fay-Herriot model.

The averages of estimates of MSE are obtained based on 5,000 simulated datasets with 1,000 replication for bootstrap, where the estimator of MSE is given in (\ref{MSEest}). 
Then the relative bias of the MSE estimator are reported in Table \ref{estmse}.
From this table, it seems that the MSE estimator gives good estimates for MSE of EBLUP.

\begin{table}[!htb]
\caption{True values of MSE of EBLUP multiplied by 100  and percentage relative gain in MSE for $m=30$, $\mu=0$, $A=1$ and $D_i$-patterns (a), (b) and (c) (values of percentage relative gain in MSE are given in parentheses).}
\begin{center}
\begin{tabular}{ccccccccccccc}
\toprule
&\multicolumn{3}{c}{Pattern (a)} & \multicolumn{3}{c}{Pattern (b)} & \multicolumn{3}{c}{Pattern (c)} \\ \midrule

$\lambda$&0.2&0.6&1.0&0.2&0.6&1.0&0.2&0.6&1.0\\ \midrule
$G_1$&12.9&14.3&16.1&12.8&14.3&15.6&12.7&14.1&15.5\\
&(13.8)&(9.5)&(6.0)&(12.9)&(7.6)&(10.1)&(12.7)&(8.3)&(6.9)\\
$G_2$&21.2&23.0&25.0&28.6&31.2&32.9&35.2&37.7&40.0\\
&(20.8)&(16.3)&(13.5)&(25.5)&(19.5)&(19.2)&(28.7)&(24.5)&(22.9)\\
$G_3$&28.4&30.6&32.7&40.3&43.4&45.7&50.0&53.0&55.8\\
&(26.4)&(20.5)&(17.8)&(34.5)&(29.1)&(27.4)&(40.2)&(36.7)&(33.3)\\
$G_4$&34.6&36.9&39.3&53.2&56.5&59.0&62.9&66.3&68.9\\
&(31.1)&(26.0)&(22.9)&(44.8)&(39.6)&(37.8)&(51.4)&(48.2)&(45.8)\\
$G_5$&39.9&42.4&45.0&59.5&63.3&65.4&71.4&74.7&77.2\\ 
&(35.4)&(29.4)&(27.5)&(50.1)&(44.8)&(43.5)&(59.0)&(55.7)&(54.0)\\
 \bottomrule
\end{tabular}
\end{center}
\label{truemse}
\end{table}

\small
\begin{table}[!htb]
\caption{Average of estimates of MSE multiplied by 100 and their relative biases for $m=30$, $\mu=0$, $A=1$ and $D_i$-patterns (a), (b) and (c) (percentage relative biases of MSE estimators are given in parentheses).}
\begin{center}
\begin{tabular}{cccccccccc}
\toprule
&\multicolumn{3}{c}{Pattern (a)} & \multicolumn{3}{c}{Pattern (b)} & \multicolumn{3}{c}{Pattern (c)} \\ \midrule

$\lambda$&0.2&0.6&1.0&0.2&0.6&1.0&0.2&0.6&1.0\\ \midrule
$G_1$&17.2&16.0&16.4&18.2&16.1&20.4&15.8&15.5&22.0\\
$G_2$ &11.9&10.1&11.8&9.9&7.2&8.9&6.7&5.6&7.8\\
$G_3$ &9.9&7.4&8.8&8.5&5.4&5.7&5.4&3.9&5.3\\
$G_4$ &8.8&6.7&6.9&7.4&4.5&4.3&5.1&3.4&4.9\\
$G_5$ &8.2&5.8&5.8&7.4&3.8&3.6&5.2&3.5&5.2\\
 \bottomrule
\end{tabular}
\end{center}
\label{estmse}
\end{table}

\normalsize

\subsection{Application to the survey data}
\label{sec:data}

We now apply the suggested procedures to the data in the Survey of Family Income and Expenditure (SFIE) in Japan.
In this study, we use the data of the spending item 'Education' in the survey in November 2011. 
The average spending (scaled by 10,000 Yen) at each capital city of 47 prefectures in Japan is obtained by $y_i$ for $i=1,\ldots,47$.
Although the average spendings in SFIE are reported every month, the sample size are around 100 for most prefectures, and data of the item 'Education' have high variability. 
On the other hand, we have data in the National Survey of Family Income and Expenditure (NSFIE) for 47 prefectures. 
Since NSFIE is based on much larger sample than SFIE, the average spendings in NSFIE are more reliable, but this survey has been implemented every five years. 
In this study, we use the data of the item 'Education' of NSFIE in 2009, which is denoted by $X_i$ for $i=1,\ldots,47$. 
Thus, we apply the dual power transformed Fay-Herriot model (\ref{model}), that is 
$$
\frac{y_i^{\lambda}-y_i^{-\lambda}}{2\lambda}=\x_i'\bbe+v_i+\varepsilon_i, \ \ \ \ i=1,\ldots,47,
$$
where $\x_i'=(1,X_i), \bbe=(\beta_1,\beta_2)'$. 
In model (\ref{model}), the variances $D_i$ are assumed to be known.
In practice, however, we need to estimate $D_i$ before applying the above model. 
In our analysis, we use the data of the spending 'Education' at the same city every November in the past ten years. 
In the usual Fay-Herriot model, we can estimate $D_i$ with the sample variance, but $D_i$ is the variance of the transformed variables in our model. 
Then, we propose an iterative method for calculating $D_i$'s. 
First we calculate the sample variance $D_i^{(0)}$'s of the log-transformed data, and we get estimates $\lah^{(0)}$ of $\la$ using $D_i^{(0)}$'s. 
Next, we recalculate the sample variance $D_i^{(1)}$'s based on the dual power transformed data with parameter $\lah^{(0)}$. 
We continue the procedure until the values of $D_i$'s converge. 
In our analysis, we get the values of $D_i$'s with 5 numbers of iterations.

We used the  REML estimators for estimation of $A$ since it performs well in simulation studies, and their estimates are $\lah=1.44$ and $\Ah=0.11$. 
The GLS estimates of $\beta_1$ and $\beta_2$ are $\beh_1=-1.09$ and $\beh_2=0.75$, so that the regression coefficient on $X_i$ is positive, namely there is a positive correlation between $y_i$ and $X_i$.
Note that the estimate of $\lambda$ is 1.44, which is far away from $0$.
This means that the logarithmic transformation does not seem appropriate for analyzing the data treated here since the treated data is not so right-skewed compared to income data. 

For model diagnostics, we calculated a correlation matrix based on the transformed data of past ten years with estimate $\lah=1.44$. 
The absolute values of each element are around $0.3$, which indicates that i.i.d assumptions of $y_1,\ldots,y_m$ is not unrealistic.
The values of EBLUP in seven prefectures around Tokyo are reported in Table \ref{emp} with the estimates of their MSEs based on $(\ref{MSEest})$. 

It is interesting to investigate what happens when one uses the log-transformed model for the same data.
When the REML estimator is used for estimation of $A$ and $\bbe$, their estimates are given by $\Ah=0.06$, $\bbeh_1=-0.90$ and $\bbeh_2=0.61$. 
Note that the estimate of $A$ in the log-transformed model is smaller than that in the dual power transformed model, which corresponds to the simulation result.  
Remember that $\Ah$ determines the rate of shrinkage of $y_i$ toward $\x_i'\bbeh$, namely, the rate increases as the value of $\Ah$ increases. 
Thus, $y_i$ in the log-transformed model are not shrunken as much as in the dual power transformed model. 
Since the dual power transformation includes the log-transformation, we can analyze positive data more flexibly with using the parametric transformed Fay-Herriot model.

\small
\begin{table}[!htb]
\caption{Values of EBLUP and their estimated MSE.}
\begin{center}
\begin{tabular}{cccccccccc}
\toprule
prefecture&$D_i$&$h(y_i,\hat{\lambda})$&$\b{x}_i'\bbeh$&$\hat{\eta}_i^{EB}$& $\widehat{{\rm MSE}}_i$\\
\midrule
Ibaraki&0.112&-0.215&-0.161&-0.188&0.075\\
Tochigi&0.444&0.002&-0.158&-0.125&0.111\\
Gunma&0.110&-0.752&-0.092&-0.429&0.073\\
Saitama&0.056&0.213&0.461&0.294&0.058\\
Chiba&0.536&1.681&0.187&0.451&0.120\\
Tokyo&0.026&0.464&0.315 &0.437&0.030\\
Kanagawa&0.188&1.068&0.235&0.551&0.097\\
 \bottomrule
\end{tabular}
\label{emp}
\end{center}
\end{table}
\normalsize

\bigskip
\noindent
{\bf Acknowledgments.}\ \

Research of the second author was supported in part by Grant-in-Aid for Scientific Research  (21540114, 23243039 and 26330036) from Japan Society for the Promotion of Science.

\medskip
\bigskip
\centerline{\bf\large{ Appendix}}

\medskip
\noindent
{\bf A.1 \ Proof of Proposition \ref{prop:1}}\ \ \  We note that the derivatives of $h^{DP}(y,\la)$ related to Assumption (A.2) are written as
\begin{align*}
h^{DP}_{\la}(y,\la)=\frac{y^{\la}+y^{-\la}}{2\la}\log y+\frac{h^{DP}(y,\la)}{\la},&
\quad
h^{DP}_y(y,\la)=\frac12(y^{\la-1}+y^{-\la-1}),
\\
h^{DP}_{y\la}(y,\la)=\frac12\log y(y^{\la-1}-y^{-\la-1}),&
\quad
h^{DP}_{\la\la}(y,\la)=h^{DP}(y,\la)(\log y)^2,
\\
\frac{d}{d\la}\Bigl(\frac{h^{DP}_{y\la}(y,\la)}{h^{DP}_y(y,\la)}\Bigr)=&\frac{4(\log y)^2}{(y^{\la}+y^{-\la})^2}.
\end{align*}
We here check whether the dual power transformation satisfies the integrability conditions in (A.3). 
Let $z (=h^{DP}(y,\la))$ be a random variable normally distributed with mean $\mu$ and variance $\sigma^2$. Then,
\begin{align*}
E[h^{DP}_{\la}(y,\la)^2]&=\frac{1}{\la^2}E\Bigl[\left( \sqrt{1+\la^2z^2}\log\bigl(\la z+\sqrt{1+\la^2z^2}\bigr)+\la z\right)^2\Bigr]\\
&<\frac{1}{\la^2}E[\left\{(1+\la^2z^2)(\la z+\la^2z^2)+\la z\right\}^2]=O(1),\\
E\left[h^{DP}(y,\la)^2h^{DP}_{\la}(y,\la)^2\right]&<\frac{1}{\la^2}E[z^2\left\{(1+\la^2z^2)(\la z+\la^2z^2)+\la z\right\}^2]=O(1),\\
\bigl|E\left\{h^{DP}_{\la\la}(y,\la)\right\}\bigr|&=\bigl|E\bigl[h^{DP}(y,\la)(\log y)^2\bigr]\bigr|={1\over \la^2} \Bigl|E\bigl[z\bigl\{\log\bigl(\la z+\sqrt{1+\la^2z^2}\bigr)\bigr\}^2\bigr]\Bigr|\\
&<E\bigl[ |z|^3(1+\la z)^2\bigr]=O(1),
\end{align*}
and
\begin{align*}
0<&E\Bigl[\frac{d}{d\la}\Bigl(\frac{h^{DP}_{y\la}(y,\la)}{h^{DP}_y(y,\la)}\Bigr)\Bigr]
=E\Bigl[\frac{4(\log y)^2}{(y^{\la}+y^{-\la})^2}\Bigr]
\\&=E\Bigl[\frac{2}{\la^2\sqrt{1+\la^2z^2}}\Bigl\{\log\bigl(\la z+\sqrt{1+\la^2z^2}\bigr)\Bigr\}^2\Bigr]<E\bigl[2z^2(1+\la z)^2\bigr]=O(1).
\end{align*}
These evaluations show that the dual power transformation satisfies (A.3). 
\hfill$\Box$

\medskip
\noindent
{\bf A.2 \ Proof of Lemma \ref{lem:1}}\ \ \ 
Since it can be easily seen that $\bbeh(\Ah(\la), \la)-\bbe=O_p(m^{-1/2})$, we here give the proof of the second part. We use $\Ah$ as abbreviation of $\Ah(\la)$ when there is no confusion.
Straightforward calculation shows that
\begin{align}
{\partial \bbeh(\Ah(\la),\la)\over \partial \la}
=\Bigl(\sum_{j=1}^m\frac{\x_j\x_j'}{\Ah+D_j}\Bigr)^{-1}\sum_{j=1}^m\frac{\x_j\x_j'\bigl(\bbeh-\bbeh^{\ast}\bigr)}{(\Ah+D_j)^2}\Bigl(\frac{\partial \Ah(\la)}{\partial \la}\Bigr)+\Bigl(\sum_{j=1}^m\frac{\x_j\x_j'}{\Ah+D_j}\Bigr)^{-1}\sum_{j=1}^m\frac{\x_jh_{\lambda}(y_j,\la)}{\Ah+D_j},  \label{dbeta}
\end{align}
where
\begin{equation}
\label{betaast}
\bbeh^{\ast}=\Bigl\{\sum_{j=1}^m\frac{\x_j\x_j'}{(\Ah+D_j)^2}\Bigr\}^{-1}\sum_{j=1}^m\frac{\x_j}{(\Ah+D_j)^2}h(y_j,\la).
\end{equation}
Since $\bbeh^{\ast}-\bbe=\O_p(m^{-1/2})$, it is seen that 
$$
\bbeh-\bbeh^{\ast}=\bbeh-\bbe-(\bbeh^{\ast}-\bbe)=\O_p(m^{-1/2}).
$$
Thus from Assumption \ref{as:2}, the expectation of the first term in (\ref{dbeta}) is $\O(m^{-1/2})$. 
For the second term in (\ref{dbeta}), we have
\begin{align*}
E&\Bigl[\Bigl(\sum_{j=1}^m\frac{\x_j\x_j'}{\Ah+D_j}\Bigr)^{-1}\sum_{j=1}^m\frac{\x_j}{\Ah+D_j}h_{\lambda}(y_j,\la)\Bigr]=\Bigl(\sum_{j=1}^m\frac{\x_j\x_j'}{A+D_j}\Bigr)^{-1}\sum_{j=1}^m\frac{\x_j}{A+D_j}E[h_{\lambda}(y_j,\la)]+\O(m^{-1/2}),
\end{align*}
where the order of the leading term of the last formula is $\O_p(1)$. Then,
\begin{align}
E[\partial \bbeh(\Ah(\la),\la)/\partial \la]
=\Bigl(\sum_{j=1}^m\frac{\x_j\x_j'}{A+D_j}\Bigr)^{-1}\sum_{j=1}^m\frac{\x_j}{A+D_j}E[h_{\lambda}(y_j,\la)]+\O(m^{-1/2}).  \label{Ebeta}
\end{align}
Therefore we obtain
\begin{align}
\notag
&\sqrt{m}\Bigl\{\partial \bbeh(\Ah(\la),\la)/\partial \la-E[\partial \bbeh(\Ah(\la),\la)/\partial \la]\Bigr\}\\
\notag
&=\Bigl(\frac1m\sum_{j=1}^m\frac{\x_j\x_j'}{A+D_j}\Bigr)^{-1}\Bigl(\frac1m\sum_{j=1}^m\frac{\x_j\x_j'}{(A+D_j)^2}\Bigr)\sqrt{m}\bigl(\bbeh-\bbeh^{\ast}\bigr)\Bigl(\frac{\partial \Ah(\la)}{\partial \la}\Bigr)\\ 
\label{dbeta2}
&+\Bigl(\frac1m\sum_{j=1}^m\frac{\x_j\x_j'}{A+D_j}\Bigr)^{-1}\frac1{\sqrt{m}}\sum_{j=1}^m\frac{\x_j}{A+D_j}\Bigl\{ h_{\lambda}(y_j,\la)-E[h_{\lambda}(y_j,\la)]\Bigr\}+\O_p(1).
\end{align}
Since $\partial \Ah(\la)/\partial \la=O_p(1)$ from (A.5) in Assumption \ref{as:2}, the first term in (\ref{dbeta2}) has $\O_p(1)$. 
For the second term in (\ref{dbeta2}), from the central limit theorem, we have
$$
\frac1{\sqrt{m}}\sum_{j=1}^m\frac{\x_j}{A+D_j}\Bigl\{h_{\lambda}(y_j,\la)-E\bigl[ h_{\lambda}(y_j,\la)\bigr] \Bigr\}=\O_p(1),
$$
which, together with Assumption \ref{as:3}, implies that the second term in (\ref{dbeta2}) is of order $\O_p(1)$. 
Therefore we can conclude that $\partial \bbeh(\Ah(\la),\la)/\partial \la-E[\partial \bbeh(\Ah(\la),\la)/\partial \la]=\O_p(m^{-1/2})$.
\hfill$\Box$

\medskip
\noindent
{\bf A.3 \ Proof of Lemma \ref{lem:2}}\ \ \ 
It is clear that condition (A.4) is satisfied for the estimators of $A$ from the results given in the literature, so that we shall verify conditions (A.5) and (A.6) in Assumption \ref{as:2}.

\ \\
{\bf PR estimator}
\medskip\noindent
For $\Ah_{PR}$ defined in (\ref{PR}), it is seen that
\begin{align}
\notag
\frac{\partial\Ah_{PR}(\la)}{\partial\la}&=\frac{2}{m-p}\sum_{j=1}^m\bigl\{h(y_j,\la)-\x_j'\bbe\bigr\}h_{\la}(y_j,\la)-\frac{2}{m-p}\sum_{j=1}^m\x_j'(\bbeh^{OLS}-\bbe)h_{\la}(y_j,\la)\\
\notag
&-\frac{2}{m-p}\sum_{j=1}^m\bigl\{h(y_j,\la)-\x_j'\bbe\bigr\}\x_j'\Bigl(\frac{\partial\bbeh^{OLS}}{\partial\la}\Bigr)+\frac{2}{m-p}\sum_{j=1}^m\x_j'(\bbeh^{OLS}-\bbe)\x_j'\Bigl(\frac{\partial\bbeh^{OLS}}{\partial\la}\Bigr),   
\end{align}
and that
$$
\frac{\partial\bbeh^{OLS}}{\partial\la}=\Bigl(\frac1m\sum_{j=1}^m\x_j\x_j'\Bigr)^{-1}\frac1m\sum_{j=1}^m\x_jh_{\la}(y_j,\la)'=O_p(1)
$$
by the law of large numbers. 
Since $\bbeh^{OLS}-\bbe=\O_p(m^{-1/2})$, we have $\partial\Ah_{PR}(\la)/\partial\la={O}_p(1)$, which shows (A.5).
For (A.6), note that 
\begin{equation}
\label{ePR}
E\Bigl[\frac{\partial\Ah_{PR}(\la)}{\partial\la}\Bigr]=\frac2{m-p}\sum_{j=1}^mE\bigl[\bigl\{h(y_j,\la)-\x_j'\bbe\bigr\}h_{\la}(y_j,\la)\bigr]+O(m^{-1/2}).
\end{equation}
Then, it is observed that
$$
\sqrt{m}\Bigl\{\frac{\partial\Ah_{PR}(\la)}{\partial\la}-E\Bigl[\frac{\partial\Ah_{PR}(\la)}{\partial\la}\Bigr] \Bigr\}=\frac2{m-p}\sum_{j=1}^mZ_j+O_p(m^{-1/2})
$$
where
\begin{equation}
\label{Zj}
Z_j=\bigl\{h(y_j,\la)-\x_j'\bbe\bigr\}h_{\la}(y_j,\la)-E\bigl[\bigl\{h(y_j,\la)-\x_j'\bbe\bigr\}h_{\la}(y_j,\la)\bigr].
\end{equation}
Since it is clear that $E[Z_j]=0$, $j=1,\ldots,m$, and $Z_1,\ldots, Z_j$ are independent, by the central limit theorem, we have
$$
\sqrt{m}\Bigl\{\frac{\partial\Ah_{PR}(\la)}{\partial\la}-E\Bigl[\frac{\partial\Ah_{PR}(\la)}{\partial\la}\Bigr] \Bigr\}=O_p(1),
$$
which shows (A.6), and Assumption \ref{as:2} is satisfied for $\Ah_{PR}$.

\ \\
{\bf FH, ML and REML estimators}

\medskip\noindent
We next show Lemma \ref{as:2} for $\Ah_{FH}, \ \Ah_{ML}$ and $\Ah_{REML}$.
For the proofs, we begin by showing that $\Ah_{FH}, \ \Ah_{ML}$ and $\Ah_{REML}$ satisfy condition (A.5).
Then we can use Lemma \ref{lem:1}, which is guaranteed under (A.4), (A.5) and Assumption \ref{as:3}.
Using Lemma \ref{lem:1}, we next show condition (A.6) for the estimators.

Since $\Ah_{FH}, \ \Ah_{ML}$ and $\Ah_{REML}$ are defined as the solutions of the equations (\ref{FH}), (\ref{ML}) and (\ref{REML}), it follows from the implicit function theorem that
\begin{equation}
\label{dA}
\frac{\partial}{\partial\la}\Ah(\la)=-\frac{G_{\la}(\la,\Ah)}{G_{A}(\la,\Ah)},
\end{equation}
where $G(\la,A)=0$ is an equation which determines an estimator of $A$, and 
$$
G_{\la}(\la,\Ah)=\frac{\partial}{\partial\la}G(\la,A)\bigg|_{A=\Ah}, \ \ \ \ \ G_{A}(\la,\Ah)=\frac{\partial}{\partial A}G(\la,A)\bigg|_{A=\Ah}
$$
For $\Ah_{FH}$, $\Ah_{ML}$ and $\Ah_{REML}$, the function $G_{\la}(\la,\Ah)$ is written as
\begin{align}
\notag
G_{\la}&(\la,\Ah)\\
\notag
&=\frac{\partial }{\partial \la}\left(\sum_{j=1}^m(A+D_j)^{-k}\bigl\{h(y_j,\la)-\x_j'\bbe(A,\la)\bigr\}^2\right)\bigg|_{A=\Ah(\la)}\\
\notag
&=2\sum_{j=1}^m(\Ah+D_j)^{-k}\bigl\{h(y_j,\la)-\x_j'\bbe\bigr\}h_{\la}(y_j,\la)-2\sum_{j=1}^m(\Ah+D_j)^{-k}\x_j'(\bbeh-\bbe)h_{\la}(y_j,\la)\\
\label{Glam}
&-2\sum_{j=1}^m(\Ah+D_j)^{-k}\bigl\{h(y_j,\la)-\x_j'\bbe\bigr\}\x_j'\bbeh_{\la}(\Ah,\la)+2\sum_{j=1}^m(\Ah+D_j)^{-k}\x_j'(\bbeh-\bbe)\x_j'\bbeh_{\la}(\Ah,\la),  
\end{align}
where
\begin{align*}
\bbeh_{\la}(\Ah,\la)&=\frac{\pd}{\pd\la}\bbe(A,\la)\bigg|_{A=\Ah}=\left(\sum_{j=1}^m\frac{\x_j\x_j'}{\Ah+D_j}\right)^{-1}\sum_{j=1}^m\frac{\x_jh_{\la}(y_j,\la)}{\Ah+D_j},
\end{align*}
which is $\O_p(1)$ under Assumptions \ref{as:2} and \ref{as:3}. Note that the case of $k=1$ corresponds to $\Ah_{FH}$, and the case of $k=2$ corresponds to $\Ah_{ML}$ and $\Ah_{REML}$. 
Using the expression of (\ref{dA}), we show that $\partial \Ah(\la)/\partial \la={O}_p(1)$, which is sufficient to verify that $G_{\la}(\la,\Ah)/m={O}_p(1)$ and $G_{A}(\la,\Ah)/m=O_p(1)$. 
For this purpose, the following facts are useful:
\begin{align}
\label{u1}
\frac{1}{m}&\sum_{j=1}^m(\Ah+D_j)^{-k}\bigl\{h(y_j,\la)-\x_j'\bbe\bigr\}h_{\la}(y_j,\la)={O}_p(1),\\
\label{u2}
\frac{1}{m}&\sum_{j=1}^m(\Ah+D_j)^{-k}\x_j'(\bbeh-\bbe)h_{\la}(y_j,\la)={O}_p(m^{-1/2}),\\
\label{u3}
\frac{1}{m}&\sum_{j=1}^m(\Ah+D_j)^{-k}\bigl\{h(y_j,\la)-\x_j'\bbe\bigr\}\x_j'=\O_p(m^{-1/2}),\\
\label{u4}
\frac1m& \sum_{j=1}^m(\Ah+D_j)^{-k}\x_j'(\bbeh-\bbe)\x_j'=\O_p(m^{-1/2}),
\end{align}
where $k=0,1,2$.
These facts can be verified by noting that $\Ah-A=O_p(m^{-1/2})$, $\bbeh-\bbe=\O_p(m^{-1/2})$ and using the law of large numbers and the central limit theorem under Assumptions \ref{as:1} and \ref{as:3}. 
If we assume that $m^{-1}G_A(\la,\Ah)=O_p(1)$ (this  is actually proved for each estimators in the end of the proof), it is immediate from (\ref{u1})$\sim$(\ref{u4}) that 
\begin{align*}
G_{\la}(\la,\Ah)/{m}=O_p(1).
\end{align*}
and we obtain $\partial \Ah(\la)/\partial \la={O}_p(1)$. 
Hence, it has been shown that  condition (A.5) is satisfied by $\Ah_{FH}$, $\Ah_{ML}$ and $\Ah_{REML}$.

We next show that condition (A.6) is satisfied by $\Ah_{FH}$, $\Ah_{ML}$ and $\Ah_{REML}$.
Since (A.4) and (A.5) are satisfied, we can use Lemma \ref{lem:1}.
Then, 
$$
\partial \bbeh(\Ah(\la),\la)/\partial \la-E\bigl[\partial \bbeh(\Ah(\la),\la)/\partial \la\bigr]=\O_p(m^{-1/2}).
$$
From (\ref{u1})$\sim$(\ref{u4}) and Lemma \ref{lem:1}, we can evaluate (\ref{Glam}) as
\begin{align*}
\frac1m G_{\la}(\la,\Ah)&=\frac2m\sum_{j=1}^m(\Ah+D_j)^{-k}\bigl\{h(y_j,\la)-\x_j'\bbe\bigr\}h_{\la}(y_j,\la)+O_p(m^{-1/2})\\
&=\frac2m\sum_{j=1}^m(A+D_j)^{-k}\bigl\{h(y_j,\la)-\x_j'\bbe\bigr\}h_{\la}(y_j,\la)+O_p(m^{-1/2})
\end{align*}
since $\Ah-A=O_p(m^{-1/2})$. 
Here we assume that 
\begin{equation}
-m^{-1}G_{A}(\la,\Ah)=c(A)+O_p(m^{-1/2}),
\label{cc}
\end{equation}
where $c(A)$ is a constant depending on $A$ with order $O(1)$.
This will be proved for each estimator in the end of this proof.
Then we have
\begin{align}
E\Bigl[\frac{\partial \Ah(\la)}{\partial \la}\Bigr]
&=E\Bigl[-\frac{m^{-1}G_{\la}(\la,\Ah)}{m^{-1}G_{A}(\la,\Ah)}\Bigr]
\label{cc0}
\\
&=c(A)^{-1}\cdot\frac2m\sum_{j=1}^m(A+D_j)^{-k}E\bigl[\bigl\{h(y_j,\la)-\x_j'\bbe\bigr\}h_{\la}(y_j,\la)\bigr]+{O}(m^{-1/2}).
\notag
\end{align}
Therefore we have
\begin{align*}
\sqrt{m}\Bigl\{\frac{\partial \Ah(\la)}{\partial \la}-E\Bigl[\frac{\partial \Ah(\la)}{\partial \la}\Bigr]\Bigr\}
&=\frac{G_{\la}(\la,\Ah)/\sqrt{m}}{G_{A}(\la,\Ah)/m}-E\Bigl[\frac{G_{\la}(\la,\Ah)/\sqrt{m}}{G_{A}(\la,\Ah)/m}\Bigr]
\\
&=c(A)^{-1}\frac2{\sqrt{m}}\sum_{j=1}^m(A+D_j)^{-k}Z_j+{O}_p(1),
\end{align*}
where $Z_j$ is given in (\ref{Zj}), and by the central limit theorem, we have
$$
\sqrt{m}\bigl[{\partial \Ah(\la)}/{\partial \la}-E\bigl\{{\partial \Ah(\la)}/{\partial \la}\bigr\}\bigr]={O}_p(1).
$$
Consequently, we have proved for $\Ah_{FH}, \Ah_{ML}$ and $\Ah_{REML}$.

It remains to show that $-m^{-1}G_{A}(\la,\Ah)=c(A)+O_p(m^{-1/2})$ for $\Ah_{FH}, \Ah_{ML}$ and $\Ah_{REML}$.
For $\Ah_{FH}$, from (\ref{FH}), we have　　　
\begin{align*}
G_A(\la,\Ah)&=-\sum_{j=1}^m(\Ah+D_j)^{-k-1}\bigl\{h(y_j,\la)-\x_j'\bbeh\bigr\}^2-2\sum_{j=1}^m(\Ah+D_j)^{-k}\bigl\{h(y_j,\la)-\x_j'\bbeh\bigr\}\x_j\Bigl(\frac{\partial }{\partial A}\bbeh(A)\Bigr),
\end{align*}
where
$$
\frac{\partial }{\partial A}\bbeh(A)=\Bigl(\sum_{j=1}^m\frac{\x_j\x_j'}{\Ah(\la)+D_j}\Bigr)^{-1}\sum_{j=1}^m\frac{\x_j\x_j'}{(\Ah(\la)+D_j)^2}\bigl(\bbeh-\bbeh^{\ast}\bigr),
$$
where $\bbeh^{\ast}$ is given in (\ref{betaast}). Note that $\bbeh-\bbeh^{\ast}=\O_p(m^{-1/2})$ and from the law of large numbers, we have
\begin{align*}
\frac{\partial }{\partial A}\bbeh(A)=\Bigl(\frac1m\sum_{j=1}^m\frac{\x_j\x_j'}{\Ah(\la)+D_j}\Bigr)^{-1}\Bigl[\frac1m\sum_{j=1}^m\frac{\x_j\x_j'}{(\Ah(\la)+D_j)^2}\Bigr]\bigl(\bbeh-\bbeh^{\ast}\bigr)=\O_p(m^{-1/2}).
\end{align*}
Thus we have
\begin{align*}
\frac1m G_A(\la,\Ah)&=-\frac1m\sum_{j=1}^m(A+D_j)^{-2}\bigl\{h(y_j,\la)-\x_j'\bbe\bigr\}^2\\
& \ \ \ \ -2\Bigl[\frac1m\sum_{j=1}^m(A+D_j)^{-1}\bigl\{h(y_j,\la)-\x_j'\bbe\bigr\}\x_j\Bigr]\Bigl(\frac{\partial }{\partial A}\bbeh(A)\Bigr)+{O}_p(m^{-1/2})\\
&=-\frac1m\sum_{j=1}^m(A+D_j)^{-2}\bigl\{h(y_j,\la)-\x_j'\bbe\bigr\}^2+{O}_p(m^{-1/2}).
\end{align*}
Since $E\bigl[\bigl\{h(y_j,\la)-\x_j'\bbe\bigr\}^2\bigr]=A+D_j$, by the law of large numbers, we have
\begin{equation}
\frac1m G_A(\la,\Ah)=-\frac1m\sum_{j=1}^m(A+D_j)^{-1}+{O}_p(m^{-1/2}),
\label{cc1}
\end{equation}
where the order of the leading term is $O(1)$, corresponding to $c(A)$.

Similarly, for $\Ah_{ML}$ and $\Ah_{REML}$ given in (\ref{ML}) and (\ref{REML}), straight calculation (almost the same as in the case of $\Ah_{FH}$) shows that   
\begin{equation}
\frac1mG_A(\la,\Ah)=-\frac1m\sum_{j=1}^m(A+D_j)^{-2}+{O}_p(m^{-1/2}),
\label{cc2}
\end{equation}
where the order of the leading term is $O(1)$, corresponding to $c(A)$.
\hfill$\Box$

\ \\
{\bf A.4 \ Proof of Lemma \ref{lem:3}}\ \ \ 
We begin by showing that $\lah-\la=O_p(m^{-1/2})$.
By the Taylor series expansion of equation (\ref{lam}), we have
$$
\hat{\la}-\la=-F(\la, \Ah, \bbeh)\Bigl(\partial F(\la, \Ah, \bbeh)/\partial \la|_{\la=\la^{\ast}}\Bigr)^{-1},
$$
where
\begin{align*}
\partial &F(\la, \Ah(\la), \bbeh(\la))/\partial \la
\\
&=\sum_{j=1}^m\frac{h_{y\la\la}(y_j,\la)}{h_y(y_j,\la)}-\sum_{j=1}^m\frac{h_{y\la}(y_j,\la) h_{y\la}(y_j,\la)}{(h_y(y_j,\la))^2}
-\sum_{j=1}^m\frac{h(y_j,\la)-\x_j'\bbeh(\Ah(\la),\la)}{\Ah(\la)+D_j}h_{\la\la}(y_j,\la)\\
&-\sum_{j=1}^m\frac{h_{\la}(y_j,\la)-\x_j'(\partial\bbeh(\Ah(\la),\la)/\partial\la)}{\Ah(\la)+D_j}h_{\la}(y_j,\la) 
+\sum_{j=1}^m\frac{h(y_j,\la)-\x_j'\bbeh(\Ah(\la),\la)}{(\Ah(\la)+D_j)^2}\frac{\partial \Ah(\la)}{\partial \la}h_{\la}(y_j,\la) 
\\
&=K_1 + K_2+K_3+K_4, \quad {\rm (say)}
\end{align*}
where $\la^{\ast}$ is satisfying $\la<\la^{\ast}<\hat{\la}$. 
For $K_1$, from Assumption \ref{as:1}, we have
$$
E\Bigl[\frac{h_{y\la\la}(y_j,\la)}{h_y(y_j,\la)}-\frac{h_{y\la}(y_j,\la)'h_{y\la}(y_j,\la)}{(h_y(y_j,\la))^2}\Bigr]=E\Bigl[\frac{\partial}{\partial\la}\Bigl(\frac{h_{y\la}(y_j,\la)}{h_y(y_j,\la)}\Bigr)\Bigr]=O(1)
$$
for $j=1,\ldots,m$. 
Since $y_1,\ldots,y_m$ are mutually independent, by the law of large numbers, we have
$$
{1\over m}K_1=\frac1m\Bigl\{\sum_{j=1}^m\frac{h_{y\la\la}(y_j,\la)}{h_y(y_j,\la)}-\sum_{j=1}^m\frac{h_{y\la}(y_j,\la)'h_{y\la}(y_j,\la)}{(h_y(y_j,\la))^2}\Bigr\}=O_p(1).
$$ 
Under Assumptions \ref{as:1} and \ref{as:2}, we have
\begin{align*}
{1\over m}K_2&=\frac1m\sum_{j=1}^m\frac{h(y_j,\la)-\x_j'\bbe}{A+D_j}h_{\la\la}(y_j,\la)-\frac1m(\Ah-A)\sum_{j=1}^m\frac{h(y_j,\la)-\x_j'\bbe}{(A^{\ast}+D_j)^2}h_{\la\la}(y_j,\la)\\
&\ \ \ -\frac1m(\bbeh-\bbe)'\sum_{j=1}^m\frac{\x_j}{A+D_j}h_{\la\la}(y_j,\la)+\frac1m(\Ah-A)(\bbeh-\bbe)'\sum_{j=1}^m\frac{\x_j}{(A^{\ast}+D_j)^2}h_{\la\la}(y_j,\la)\\
&=\frac1m\sum_{j=1}^m\frac{h(y_j,\la)-\x_j'\bbe}{A+D_j}h_{\la\la}(y_j,\la)+O_p(m^{-1/2})=O_p(1).
\end{align*}
Similarly, we can evaluate $K_3$ as
\begin{align*}
\frac1m K_3=&\frac1m\sum_{j=1}^m\frac{h_{\la}(y_j,\la)-\x_j'(\partial\bbeh(\Ah(\la),\la)/\partial\la)}{A+D_j}h_{\la}(y_j,\la)\\
&\ \ \ \ -\frac{(\Ah-A)}{m}\sum_{j=1}^m\frac{h_{\la}(y_j,\la)-\x_j'(\partial\bbeh(\Ah(\la),\la)/\partial\la)}{(A^{\ast}+D_j)^2}h_{\la}(y_j,\la),
\end{align*}
which is of order $O_p(1)$ under Assumptions \ref{as:1} and \ref{as:2}.
Moreover, 
\begin{align*}
\frac1m K_4&=\frac1m\Bigl(\frac{\partial \Ah(\la)}{\partial \la}\Bigr)\sum_{j=1}^m\frac{h(y_j,\la)-\x_j'\bbe}{(A+D_j)^2}h_{\la}(y_j,\la)
-\frac{(\Ah-A)}{m}\Bigl(\frac{\partial \Ah(\la)}{\partial \la}\Bigr)\sum_{j=1}^m\frac{h(y_j,\la)-\x_j'\bbe}{2(A^{\ast}+D_j)^3}h_{\la}(y_j,\la)\\
&\ \ \ \ -\frac1m\Bigl(\frac{\partial \Ah(\la)}{\partial \la}\Bigr)\sum_{j=1}^m\frac{(\bbeh-\bbe)'\x_j}{(A+D_j)^2}h_{\la}(y_j,\la)
+\frac{(\Ah-A)}{m}\Bigl(\frac{\partial \Ah(\la)}{\partial \la}\Bigr)\sum_{j=1}^m\frac{(\bbeh-\bbe)'\x_j}{2(A^{\ast}+D_j)^3}h_{\la}(y_j,\la),
\end{align*}
which is of order $O_p(1)$.
As a result, we have
$$
\frac1m\Bigl\{\frac{\partial F(\la, \Ah(\la), \bbeh(\la))}{\partial \la}\bigg|_{\la=\la^{\ast}}\Bigr\}=O_p(1).
$$
Furthermore, by Assumption \ref{as:1}, we have
\begin{align*}
F&(\la, \Ah(\la), \bbeh(\la))\\
&=\sum_{j=1}^m\frac{h_{y\la}(y_j,\la)}{h_y(y_j,\la)}-\sum_{j=1}^m\frac{h(y_j,\la)-\x_j'\bbeh(\Ah(\la),\la)}{\Ah(\la)+D_j}h_{\la}(y_j,\la)\\
&=\sum_{j=1}^m\frac{h_{y\la}(y_j,\la)}{h_y(y_j,\la)}-\sum_{j=1}^m\frac{h(y_j,\la)-\x_j'\bbe}{A+D_j}h_{\la}(y_j,\la)-(\Ah-A)\sum_{j=1}^m\frac{h(y_j,\la)-\x_j'\bbe}{(A^{\ast}+D_j)^2}h_{\la}(y_j,\la)\\
&\ \ \ -(\bbeh-\bbe)'\sum_{j=1}^m\frac{\x_j}{A+D_j}h_{\la}(y_j,\la)+(\Ah-A)(\bbeh-\bbe)'\sum_{j=1}^m\frac{\x_j}{(A^{\ast}+D_j)^2}h_{\la}(y_j,\la),
\end{align*}
which is evaluated as
$$
\sum_{j=1}^m\Bigl\{\frac{h_{y\la}(y_j,\la)}{h_y(y_j,\la)}-\frac{h(y_j,\la)-\x_j'\bbe}{A+D_j}h_{\la}(y_j,\la)\Bigr\}+O_p(m^{1/2}).
$$
For all $j=1,\ldots,m$, we have
$$
E\Bigl[\frac{h_{y\la}(y_j,\la)}{h_y(y_j,\la)}-(A+D_j)^{-1}\bigl\{h(y_j,\la)-\x_j\bbe\bigr\}h_{\la}(y_j,\la)\Bigr]
=E\Bigl[\frac{\partial\log f(Y_j; \la, \bbe,A)}{\partial\la}\Bigr]=0,
$$
where $f(y_j; \la, \bbe,A)$ is the density function of observation $y_j$ in (\ref{model}). 
By the central limit theorem, we have
$$
\frac{1}{\sqrt{m}}F(\la, \Ah(\la), \bbeh(\la))=O_p(1).
$$
Therefore we have
$$
\sqrt{m}(\hat{\la}-\la)=-\frac{1}{\sqrt{m}}F(\la, \Ah(\la), \bbeh(\la))\Bigl\{\frac1m\Bigl(\partial F(\la, \Ah, \bbeh)/\partial \la|_{\la=\la^{\ast}}\Bigr)\Bigr\}^{-1}=O_p(1),
$$ 
and we conclude that $\hat{\la}-\la=O_p(m^{-1/2})$.

\medskip
We next show that $E[\lah-\la]=O(m^{-1})$.
From the first part of Lemma \ref{lem:3}, we have $\hat{\la}-\la=O_p(m^{-1/2})$. 
Then expanding (\ref{lam}) shows that
$$
\hat{\la}-\la=-F(\la, \Ah, \bbeh)\bigl(\partial F(\la, \Ah, \bbeh)/\partial \la\bigr)^{-1}+O_p(m^{-1}).
$$
Thus, it is sufficient to show that the expectation of the first term is $O(1/m)$. 
It is observed that 
\begin{align*}
E\Bigl[F(\la, \Ah, \bbeh)\Bigl\{\Bigl(\partial F(\la, \Ah, \bbeh)/\partial \la\Bigr)\Bigr\}^{-1}\Bigr]&=E\Bigl[\Bigl\{\frac1mF(\la, \Ah, \bbeh)\Bigr\}\Bigl\{\frac1mE\bigl[\partial F(\la, \Ah, \bbeh)/\partial \la\bigr]+O_p(m^{-1/2})\Bigr\}^{-1}\Bigr]\\
&=E\Bigl[\frac1mF(\la, \Ah, \bbeh)\Bigr]\Bigl\{\frac1mE\bigl[\partial F(\la, \Ah, \bbeh)/\partial \la\bigr]\Bigr\}^{-1}+O(m^{-1}).
\end{align*}
Since $E[\Ah-A]=O(m^{-1})$ and $E[\bbeh-\bbe]=\O(m^{-1})$, it is noted that
\begin{align*}
E\Bigl[\frac1mF(\la, \Ah, \bbeh)\Bigr]
&=-E(\Ah-A)\cdot\frac1m\sum_{j=1}^m\frac{E\bigl[\bigl\{h(y_j,\la)-\x_j'\bbe\bigr\}h_{\la}(y_j,\la)\bigr]}{(A^{\ast}+D_j)^2}\\
&\ \ \ -E(\bbeh-\bbe)'\cdot\frac1m\sum_{j=1}^m\frac{\x_j}{A+D_j}E\bigl[h_{\la}(y_j,\la)\bigr]+O(m^{-1}),
\end{align*}
which is of order $O(m^{-1})$.
Hence,
$$
E\Bigl[F(\la, \Ah, \bbeh)\Bigl\{\Bigl(\partial F(\la, \Ah, \bbeh)/\partial \la\Bigr)\Bigr\}^{-1}\Bigr] = O(m^{-1}).
$$
Since 
\begin{align*}
\frac1mE\bigl[\partial F(\la, \Ah, \bbeh)/\partial \la\bigr]=O(1),
\end{align*}
it is cocluded that $E[\hat{\la}-\la]=O(m^{-1})$.
\hfill$\Box$

\ \\
{\bf A.5 \ Proof of Lemma \ref{lem:5}}\ \ \ \
By the Taylor series expansion of $\hat{\eta_i}^{EB1}$, we have
\begin{align*}
\hat{\eta_i}^{EB1}-\hat{\eta_i}^{B}&=\frac{D_i}{A+D_i}\x_i'(\bbeh-\bbe)+\frac{D_i}{(A+D_i)^2}(\Ah-A)\bigl\{h(y_i,\la)-\x_i'\bbe\bigr\}\\
&\ \ \ -\frac{D_i}{(A^{\ast}+D_i)^2}\x_i'(\Ah-A)(\bbeh-\bbe)
-\frac{D_i}{(A^{\ast}+D_i)^3}\bigl\{h(y_i,\la)-\x_i'\bbe^{\ast}\bigr\}(\Ah-A)^2,
\end{align*}
where $A^{\ast}$ is an intermediate value of $A$ and $\Ah$ and $\bbe^{\ast}$ is an intermediate vector of $\bbe$ and $\bbeh$. Differentiating the both sides by $\la$, we have
\begin{align*}
\frac{\partial }{\partial \la}\hat{\eta_i}^{EB1}&=\frac{\partial }{\partial \la}\hat{\eta_i}^{B}
+\frac{D_i}{A+D_i}\x_i'\Bigl(\frac{\partial }{\partial \la}\bbeh(\la)\Bigr)+\frac{D_i}{(A+D_i)^2}\Bigl(\frac{\partial }{\partial \la}\Ah(\la)\Bigr)\bigl\{h(y_i,\la)-\x_i'\bbe\bigr\}\\
&\ \ \ -\frac{D_i}{2(A^{\ast}+D_i)^2}\x_i'\Bigl(\frac{\partial }{\partial \la}\Ah(\la)\Bigr)(\bbeh-\bbe)
-\frac{D_i}{2(A^{\ast}+D_i)^2}\x_i'(\Ah-A)\Bigl(\frac{\partial }{\partial \la}\bbeh(\la)\Bigr)\\
&\ \ \ -\frac{2D_i}{(A^{\ast}+D_i)^3}h_{\la}(y_i,\la)(\Ah-A)\Bigl(\frac{\partial }{\partial \la}\Ah(\la)\Bigr)\\
&=\frac{\partial }{\partial \la}\hat{\eta_i}^{B}+\frac{D_i}{(A+D_i)^2}E\Bigl[\frac{\partial }{\partial \la}\Ah(\la)\Bigr]\bigl\{h(y_i,\la)-\x_i'\bbe\bigr\}+\frac{D_i}{A+D_i}\x_i'E\Bigl[\frac{\partial }{\partial \la}\bbeh(\la)\Bigr]+O_p(m^{-1/2}),
\end{align*}
from Lemmas \ref{lem:1} and \ref{lem:2}. 
Also from Lemmas \ref{lem:1} and \ref{lem:2}, we already know that
\begin{equation}
E\Bigl[\frac{\partial }{\partial \la}\bbeh(\la)\Bigr]=\Bigl(\sum_{j=1}^m\frac{\x_j\x_j'}{A+D_j}\Bigr)^{-1}\sum_{j=1}^m\frac{\x_j}{A+D_j}E\bigl[h_{\la}(y_j,\la)\bigr]+\O(m^{-1/2}),
\label{22}
\end{equation}
and
\begin{equation}
E\Bigl[\frac{\partial }{\partial \la}\Ah(\la)\Bigr]=\Bigl(\sum_{j=1}^m(A+D_j)^{-k}\Bigr)^{-1}\Bigl(\sum_{j=1}^m\frac{E[\{h(y_j,\la)-\x_j'\bbe\}h_{\la}(y_j,\la)]}{(A+D_j)^{k}}\Bigr)+O(m^{-1/2}),
\label{23}
\end{equation}
where $k=1$ corresponds to $\Ah_{FH}$ and $k=2$ corresponds to $\Ah_{ML}$ and $\Ah_{REML}$. 
The formula (\ref{22}) comes from (\ref{Ebeta}), and the formula (\ref{23}) is obtained by combining (\ref{cc}), (\ref{cc1}) and (\ref{cc2}).
For $\Ah_{PR}$, from (\ref{ePR}),
$$
E\Bigl[\frac{\partial \Ah_{PR}(\la)}{\partial \la}\Bigr]=\frac2{m-p}\sum_{j=1}^mE\bigl[\bigl\{h(y_j,\la)-\x_j'\bbe\bigr\}h_{\la}(y_j,\la)\bigr]+O(m^{-1/2}),
$$
which completes the proof.
\hfill$\Box$

\ \\

\end{document}